\documentclass[12pt, a4paper]{article}
\pdfoutput=1
\usepackage{jheppub}
\usepackage[toc,page]{appendix}
\usepackage[strict]{changepage}

\usepackage[export]{adjustbox}
\usepackage{hyperref}
\usepackage{color}
\usepackage{graphicx}
\usepackage{color}
\usepackage{overpic}
\usepackage{float}
\usepackage[mathscr]{euscript}
\usepackage[usenames,dvipsnames]{xcolor}
\usepackage{amsmath,amsfonts,amssymb,verbatim,float,mathtools,bbold}
\usepackage{physics}
\usepackage{resizegather}
\usepackage{tikz}
\usetikzlibrary{shapes.misc}
 \tikzset{cross/.style={cross out, draw=black, minimum size=2*(#1-\pgflinewidth), inner sep=0pt, outer sep=0pt},
cross/.default={4pt}}

\usepackage[most]{tcolorbox}
\tcbset{myformula/.style={colback=white, 
    colframe=black, 
    top=4pt,bottom=4pt,left=0pt,right=4pt,
    boxsep=0pt,
    arc=0pt,
    outer arc=0pt,
    fit algorithm=hybrid*
}}
\newcommand\fiteq[1]{%
  \sbox{\mybox}{$\displaystyle#1$}%
  \ifdim\wd\mybox>.85\textwidth\resizebox{.85\textwidth}{!}{\usebox{\mybox}}%
  \else\usebox{\mybox}\fi%
}
\newsavebox{\mybox}

\newtcolorbox{equationframe}{
math
}

\newcommand{\quarter}{{1\over 4}}
\newcommand{\half}{{1\over 2}}
\newcommand{\OO}{\mathcal{O}}
\newcommand{\JJ}{\mathcal{J}}

\newcommand{\sgn}{{\rm sgn}}
\newcommand{\rhoS}[1]{\rho^{#1 }}
\def\sign{ {\rm sgn} }

\newcommand{\Calpha}{{C(t;\beta)_{(\alpha,0)}}}
\newcommand{\CalphaOm}{{C(\omega;\beta)_{(\alpha,0)}}}
\newcommand{\rhoh}{\rho^{{1\over 2}}}
\newcommand{\rhoq}{\rho^{{1\over 4}}}

\def\half{{1 \over 2}}

\def\k{{\vec{k}}}

\def\p{{\vec{p}}}

\def\CO{\mathcal{O}}

\def\CA{{\cal A}}
\def\CF{{\cal F}}

\def \k {{\vec{k}}}
\def\Or[#1]{{\text{O}}\left({#1}\right)}
\def\dotl[#1,#2]{\left\langle #1,\, #2 \right\rangle}
\def\dotlb[#1,#2]{\left\langle #1,\, #2 \right\rangle}
\def\dotlm[#1,#2]{\left[ #1,\, #2 \right]}
\def\dotp[#1,#2]{(\vect{#1} \cdot\vect{#2})}
\def\aff[#1,#2]{\hat{#1}(#2)}
\def\n4sym{{\cal N}=4 SYM}
\def\>{\rangle}
\def\<{\langle}
\def\({\left(}
\def\){\right)}
\def\weight[#1,#2,#3]{\{(#1),#2,#3\}}
\def\ads[#1]{$\text{AdS}_{#1}$}

\newcommand{\be}{\begin{equation}}
\newcommand{\ee}{\end{equation}}
\newcommand{\beq}{\begin{eqnarray}}
\newcommand{\eeq}{\end{eqnarray}}
\newcommand{\ba}{\begin{align}}
\newcommand{\ea}{\end{align}}

\renewcommand{\vec}[1]{\boldsymbol{#1}}
\newcommand{\bs}{\begin{split}}
\def\sess\end{split}

\newcommand{\vect}[1]{{\boldsymbol{#1}}}

\definecolor{col1}{RGB}{153, 52, 121}

\makeatletter
\def\@fpheader{\relax}
\makeatother
\begin{document}

\title{Regularization dependence of the OTOC. Which Lyapunov spectrum is
  the physical one?}
\author{Aurelio Romero-Berm\'udez,~}
\author{Koenraad Schalm,~}
\author{Vincenzo Scopelliti}
\affiliation{Instituut-Lorentz, $\Delta$ITP, Universiteit Leiden, P.O. Box 9506, 2300 RA Leiden, The Netherlands}

\emailAdd{romero, kschalm, scopelliti@lorentz.leidenuniv.nl}
\abstract{We study the contour dependence of the out-of-time-ordered
  correlation function  (OTOC)  both in weakly coupled field theory
  and in the Sachdev-Ye-Kitaev (SYK) model. We show that its value,
  including its Lyapunov spectrum, depends sensitively on the shape of
  the complex time contour in generic weakly coupled field
    theories. For gapless theories with no thermal mass, such as SYK, the Lyapunov spectrum turns out to be an
    exception; their Lyapunov spectra do not exhibit contour
    dependence, though the full OTOCs do.  Our result puts into question
  which of the Lyapunov exponents computed from the exponential
  growth of the OTOC reflects the actual physical dynamics of the
  system. We argue that, in a weakly coupled $\Phi^4$ theory, a
  kinetic theory argument indicates that the symmetric configuration of the time
  contour, namely the one for which the bound on chaos has been
  proven, has a proper interpretation in terms of
  dynamical chaos. Finally, we point
out that a relation between these OTOCs and a quantity which may be
measured experimentally --- the Loschmidt echo --- also suggests a
symmetric contour configuration, with the subtlety that the inverse periodicity
in Euclidean time is half the physical temperature. In this
interpretation the chaos bound reads \mbox{$\lambda \leq \frac{2\pi}{\beta}
= \pi T_{\text{physical}}$}.
 }

\maketitle

\section{Introduction}
It has long been known that chaos, understood as the exponential
sensitivity of the dynamics to initial conditions, does not have an 
immediate equivalent
in the quantum dynamics governed by the Schr\"odinger equation. In
quantum systems one
needs to define quantum chaos in a more indirect way. One way to do
so, is to measure the correlation between an operator $W(t)$ and
some earlier perturbation $V(0)$ and compare this with the correlation 
where the perturbation $V(0)$ is performed after operator $W(t)$ is inserted:
\begin{align}
  \label{eq:1}
  \langle \psi_{\rm final}|W(t)V(0)|\psi_{\rm initial}\rangle - \langle
  \psi_{\rm final}|V(0)W(t)|\psi_{\rm initial}\rangle = \langle \psi_{\rm final}|[W(t),V(0)]|\psi_{\rm initial}\rangle \,.
\end{align}
Choosing $W(t) = q(t)$ and $V(0)=p(0)$ this commutator formally equals
$[W(t),V(0)] = i\hbar\frac{\partial q(t)}{\partial q(0)}$ and in that
sense the above measures the sensitivity to initial conditions. The
commutator is evaluated between two wave-functions, however. 
For a generic $|\psi_{\rm initial}\rangle$ and $|\psi_{\rm   final}\rangle$, this is a complex amplitude that also depends on the details of both. 
An obvious step is to sum over final states, which converts this to an expectation value
\begin{align}
 C(t;\psi_{\rm  initial}) &=   \sum_{\rm  final}
 \langle \psi_{\rm  initial}|[W(t),V(0)]^{\dagger}|\psi_{\rm final}\rangle 
 \langle  \psi_{\rm  final}|[W(t),V(0)]|\psi_{\rm initial}\rangle\nonumber\\
 &= 
\langle \psi_{\rm  initial}|[W(t),V(0)]^{\dagger}[W(t),V(0)]|\psi_{\rm initial}\rangle\,.  \label{eq:1b}
\end{align}
To also isolate the dynamics driven by $V(0)$ and $W(t)$ as much from the
details of the initial state, one can average over a
suitable ensemble. A physically natural choice is the thermal one
\begin{align}
  \label{eq:2}
  C(t;\beta) 
 &= \sum_{\rm initial} e^{-\beta E[\psi_{\rm initial}]}
 \langle \psi_{\rm initial}|[W(t),V(0)]^{\dagger}[W(t),V(0)]|\psi_{\rm initial}\rangle
  \nonumber \\
&= \Tr\rho_{\beta}[W(t),V(0)]^{\dagger}[W(t),V(0)] \,.
\end{align}
This commutator-squared $C(t;\beta)$ or, equivalently, this out-of-time ordered
correlation function (OTOC) has been of much interest as a diagnostic
of chaotic behaviour in many-body systems
\cite{Shenker2014a,Shenker2014,Maldacena2016c}. Specifically, if this OTOC has a
regime where it exhibits an
exponential time dependence, $C(t) \sim e^{\lambda t}$, this behaviour
has been proposed to be a signature of chaos, with  $\lambda$ being
the \textit{quantum} Lyapunov exponent.\footnote{Note that the
  Lyupanov exponent defined this way is in fact twice the chaos
  exponent one would surmise from the choice $W(t)=q(t)$, $V(0)=p(0)$
  with $q(t) \sim e^{\lambda_{\text{chaos}}t}q(0)$, i.e. $\lambda=2\lambda_{\text{chaos}}$.} Moreover, this quantum Lyapunov exponent has been conjectured to be bounded from above
$\lambda \leq 2\pi k_B T/\hbar$
 \cite{Maldacena2016c}.

In practice most computations do not compute $C(t)$ as defined
above. Rather one ``smears'' the thermal distribution between the two
commutators \cite{Maldacena2016c,Chowdhury2017}
\begin{align}
  \label{eq:4}
  C(t;\beta)_{\text{regulated}}\equiv \Tr\left(\rhoh [W(t),V]^\dagger\rhoh [W(t),V]\right)\,.
\end{align}
Mathematically, this has the advantage of being manifestly Hermitian
(see e.g. \cite{Chowdhury2017}). The physical intuition is that in a
QFT this correlation function naively suffers from a short-distance
divergences caused by the insertion of two operators at the same
time. As chaos is in principle a long-time characteristic, the claim is that the information about chaos, and in particular
the Lyapunov exponents $\lambda$, do not depend on this regularization
\cite{Maldacena2016c,Chowdhury2017}.

We will show that this intuition is incorrect, as was also pointed
out earlier in \cite{Liao:2018uxa} for the specific case of 2D fermions with quenched disorder. By explicit computation
we will show that in the two-parameter family of ``regularized'' OTOCs
\begin{align}
\label{eq:5b}
&C(t;\beta)_{(\alpha,\sigma)}\equiv  \\ 
&
-\Tr\left[
\rhoS{1-\alpha-\sigma} \Big(W^{\dagger}(t) \rhoS{\sigma}V^{\dagger}-V^{\dagger} \rhoS{\sigma}W^{\dagger}(t)\Big)
 \rhoS{\alpha-\sigma} \Big(W(t) \rhoS{\sigma}V-V \rhoS{\sigma}W(t)\Big)\right]
 \nonumber \,,
\end{align}
the Lyapunov exponents are independent of $\sigma$ but do depend on
$\alpha$. 
Our computation shows that this regularization 
dependence is an IR-effect and has nothing to
do with short-distance singularities. The more appropriate comparison
for the regularization dependence of the OTOC
is the proof in Schwinger-Keldysh theory that physical correlation
functions are independent on the choice of contour. In
Schwinger-Keldysh theory, there is a diagrammatic
proof that \textit{physical} Green's functions  involving operator
insertions either on only forward or only backward branches are independent of
the contour due to energy conservation; this can be found in
e.g. \cite{Matsumoto1983,Matsumoto1984}. The OTOC, however, is a
correlation on a doubled Schwinger-Keldysh contour \cite{Stanford2016}
and the two-body Green's functions involved in the commutator-squared
involve operators inserted on both forward and backward
branches. The arguments of
\cite{Matsumoto1983,Matsumoto1984} do not generalize to prove that the
correlation functions that appear in $C(t;\beta)_{(\alpha,\sigma)}$ must be independent on the
contour. Our explicit computation in Section \ref{sec:weak} shows that they indeed are not.

Gapless theories are notoriously more IR sensitive than gapped
  theories. Perhaps somewhat counterintuitively, our results show that
  weakly coupled gapless theories are in fact less contour-dependent than
  explicitly gapped theories, although the thermally generated mass
  does imbue a suppressed dependence. The SYK model on the other hand,
which has been at the forefront of many OTOC studies, has no thermally
induced mass. In this model specifically the contour dependence is
extremely weak. In fact its Lyapunov spectrum turns
out to be always contour-independent, as we show in Section \ref{sec:SYK}.

Let us stress that the found contour dependence in generic models is not a pedantic point. As also pointed out by \cite{Liao:2018uxa}, OTOCs are now being measured either in
numerical or actual physical experiments. Often one massages the
regulator to be the most convenient for the set-up. For instance, Das
et. al. \cite{Das2018} use the canonical thermal OTOC $C(t;\beta)_{(0,0)}$ in a numerical study, whereas a cold atom experiment measures a Loschmidt echo
\cite{Garttner2017}, which can be related to
$C(t;\beta)_{(\frac{1}{2},0)}$. As the theoretical prediction for these two correlation
functions is different due to the regulator dependence, these two
experimental results cannot be compared to each other. 

Given the regularization dependence that we and \cite{Liao:2018uxa}
observe, the immediate question arises: which is the proper regularization that measures
quantum chaos. As the previous paragraph shows, to some extent this is in
the eye of the beholder. One can devise experimental set-ups that
measure either. Nevertheless, we will argue that the OTOC that most
closely reflects physical microscopic chaos is the symmetrized one
$C(t;\beta)_{(\frac{1}{2},0)}$ used originally for hermiticity reasons. Our
  argument rests on the following fact: in weakly coupled field theories the computation of any of
    the OTOCs $C(t;\beta)_{(\alpha,\sigma)}$ can be cast in the form of a kinetic
    equation \cite{Grozdanov:2018atb}. This kinetic equation reveals most closely the physical
    process one is actually computing. In terms of the kinetic
    equation, only the
    symmetrized OTOC with $\alpha=1/2$ can be understood as a microscopic
    unbiased ``collision''-counter. Such unbiased collision counters
    have long been successfully proposed as tracking microscopic
    classical chaos
    \cite{PhysRevLett.80.2035, VanZon2000}. This is explained in Section  \ref{sec:kinet-theory-interpr}.

We conclude by showing the symmetric OTOC $C(t;\beta)_{(\frac{1}{2},0)}$ 
regulated this way has a natural
interpretation as a Loschmidt echo, rather than an expectation value
in a thermal ensemble as in the introductory thought experiment. 
This has as subtle physical consequence that
the physical temperature is set by twice the inverse periodicity in
Euclidean time. In this interpretation the MSS bound reads
\begin{align}
  \label{eq:6}
  \lambda \leq {k_B\over \hbar}\frac{2\pi}{\beta} ={ \pi k_B T_{\text{physical}}\over \hbar}~.
\end{align}
   
\section{A two-parameter family of extended Schwinger-Keldysh contours}  \label{sec:regularizations}
  
We will assume that $W(t)$ and $V(0)$ are hermitian from here on. 

\noindent
We formally consider the following regularization of the commutator-squared of  Eq. \eqref{eq:2}:
\begin{align}
&C(t;\beta)_{(\alpha,\sigma)}=
\Tr\left[{\CA^\dagger\CA}
\right]\geq0\,,\\
&\CA\equiv  \rho^{\alpha-\sigma\over 2}\ [W(t),V(0)]_{\sigma} \ \rho^{1-\alpha-\sigma\over 2}\,, \quad [A,B]_{\sigma}\equiv A\rho^{\sigma} B-B\rho^{\sigma} A\,,
\end{align}
with  $\sigma\in[0,1/4]$.  First, we note that for
$0\leq\alpha\leq1$, $\Calpha$ is positive definite and for
$\alpha=\{0,1\}$, $\sigma=0$ we recover the unregulated thermal commutator-squaredd in the thermal state.

Expanding the terms in
$C(t;\beta)_{(\alpha,\sigma)}$  gives Eq.~\eqref{eq:5b}
\begin{align}
C(t;\beta)_{(\alpha,\sigma)}&=
-\Tr\left[
\rhoS{1-\alpha-\sigma} W(t) \rhoS{\sigma}V
 \rhoS{\alpha-\sigma} W(t)\rhoS{\sigma}V
+
\rhoS{1-\alpha-\sigma} V\rhoS{\sigma}W(t) 
\rhoS{\alpha-\sigma} V\rhoS{\sigma}W(t)
\right]\nonumber\\
&\phantom{=}\ +
\Tr\left[
\rhoS{1-\alpha-\sigma} W(t) \rhoS{\sigma}V 
\rhoS{\alpha-\sigma} V\rhoS{\sigma}W(t)
+
\rhoS{1-\alpha-\sigma} V\rhoS{\sigma}W(t) 
\rhoS{\alpha-\sigma} W(t)\rhoS{\sigma}V
\right]\,,\nonumber
\end{align}
The last two are conventional Schwinger-Keldysh time-ordered
correlation functions (TOCs), whereas the first two are true
out-of-time-ordered correlators of the type 
\begin{align}
F(t_1,t_2)_{(\alpha,\sigma)} &\equiv \Tr[\rhoS{1-\alpha-\sigma}W(t_1)\rhoS{\sigma}V\rhoS{\alpha-\sigma} W(t_2)\rho^{\sigma} V]\notag\\
&=\Tr[\rhoS{1-\alpha}\ W(t_1-i\sigma\beta) V\  \rhoS{\alpha} \ W(t_2-i\sigma\beta) V]\notag\\
&=F(t_1-i\sigma\beta,t_2-i\sigma\beta)_{(\alpha,0)}\,.
\end{align}
Schematically $C(t;\beta)_{(\alpha,\sigma)} $ equals
\begin{align}
  C(t;\beta)_{(\alpha,\sigma)}  &= \text{TOCs}{-}
  F\Big(t{-}i\sigma\beta,t{-}i\sigma\beta\Big)_{(\alpha,0)}{-}
F\Big(t{-}i(1{-}\alpha{-}\sigma)\beta,t{-}i(\alpha{-}\sigma)\beta\Big)_{(\alpha,0)} \,.  \label{eq:8}
\end{align}
Each out-of-time ordered correlator $F(t_1,t_2)_{(\alpha,\sigma)}$
may
be seen as a correlation function in the extended Schwinger-Keldysh
contour. The usual choice with $\alpha=1/2$, $\sigma=1/4$ is shown in
Fig. \ref{fig:SK_sigma}-(a); the more general
$F(t_1,t_2)_{(\alpha,\sigma)}$ corresponds to a more complicated
contour like the one shown in Fig.  \ref{fig:SK_sigma}-(b) with
different separations in imaginary time between each of the branches.

It is this OTOC $F(t_1,t_2)_{(\alpha,\sigma)}$ that controls the regime
of exponential growth and the Lyapunov spectrum
$F(t_1,t_2)_{(\alpha,\sigma)} \sim 1-
\mathsf{A}(t_1{-}t_2) e^{\lambda_{(\alpha,\sigma)}\frac{(t_1+t_2)}{2}} $,
with $\mathsf{A}(0)$ a finite positive number.
We will now show that
the same exponential time dependence and thus the same Lyapunov
exponent is obtained independent of the value of $\sigma$ if $\alpha=1/2$.
This follows directly  from the analyticity property of the function
highlighted above:
$F(t_1,t_2)_{(\alpha,\sigma)} =
F(t_1-i\sigma\beta,t_2-i\sigma\beta)_{(\alpha,0)}$  for $ 0\leq \sigma
\leq \text{min}\{\alpha,1-\alpha\}$ 
\cite{Maldacena2016c}. Suppose for the particular value $\sigma =0$ the function
$F(t_1,t_2)_{(\alpha,0)}$ has the exponential behavior
$F(t_1,t_2)_{(\alpha,0)}\simeq  \mathsf{A}(t_1{-}t_2)
e^{\lambda_\alpha\frac{t_1+t_2}{2}}$ with $\lambda_\alpha=\lambda_{(\alpha,0)}$. Analyticity
implies that
$F(t_1{+}i\xi\beta,t_2{+}i\gamma\beta)_{\left(\alpha,0\right)}\simeq
 \mathsf{A}(t_1{-}t_2+i\beta(\gamma{-}\xi~\text{mod} 1))
e^{i{\xi+\gamma\over2}\beta\lambda_\alpha}e^{\lambda_\alpha
  \frac{t_1+t_2}{2}}$. 
Substituting this into Eq. \eqref{eq:8}, we get
\begin{align}
  C(t;\beta)_{(\alpha,\sigma)} 
 &{\simeq}\text{TOCs}{ +}\Big[\mathsf{A}(0) 
  e^{-i\sigma\beta\lambda_\alpha} 
  {+} \mathsf{A}\big(i\beta(1{-}2\alpha~\text{mod} 1)\big)
   e^{-i{\beta\lambda_\alpha\over2}(1-\alpha-\sigma+\alpha-\sigma)}\Big]
   e^{\lambda_{\alpha}t}~.
\end{align}
For the specific choice $\alpha =1/2$ --- the one that is made in almost all
previous studies --- the prefactor $\mathsf{A}(i\beta(1{-}2\alpha~\text{mod} 1))|_{\alpha=1/2} = \mathsf{A}(0)$ is the same in both cases and equal to
the one computed for the $\alpha=1/2$. Thus
\begin{align}
 C(t;\beta)_{(\frac{1}{2},\sigma)}   &{=}\text{TOCs}{ +}2\mathsf{A}(0)e^{-i\beta{\lambda_{1/2}\over4}}\cos\left[\left(\sigma-\quarter\right)\beta\lambda_{1/2}\right]e^{\lambda_{1/2} t}\,,\label{eq:Csigma}
\end{align}
with $\lambda_{1/2}=\lambda_{({\frac{1}{2}},\sigma)},\, \forall \sigma$.
Although the Lyapunov exponent is not affected by the 
deformation parametrized by $\sigma$ away from $(\alpha,\sigma)=(\frac{1}{2},0)$, we do
see that the prefactor of the exponential depends on the
$\sigma$-deformation of the contour. Therefore,  similarly to a Wightman function in
Schwinger-Keldysh theory, the full commutator-squared
$C(t;\beta)_{(\frac{1}{2},\sigma)}$ cannot be an observable measurable in
an experiment, even though it may contain  physical information.

\begin{center}
 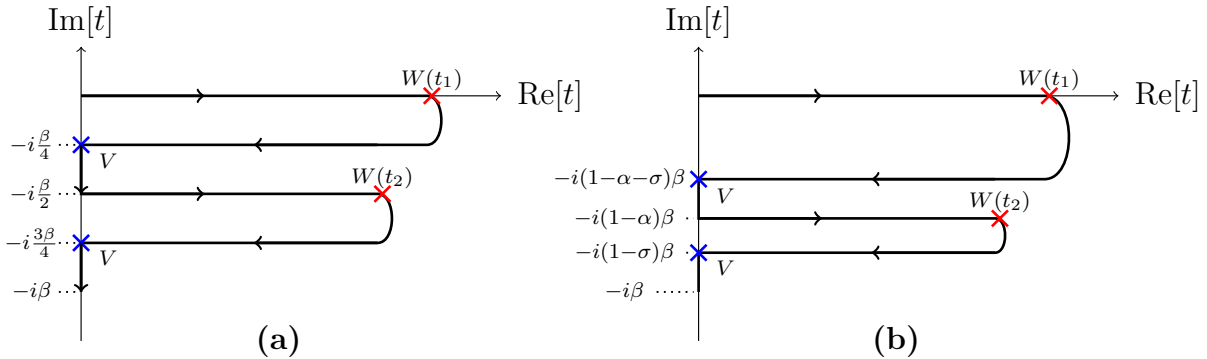
\begin{figure}[t]
\begin{tikzpicture}[scale=0.65][h]
\hspace{-5mm}
\begin{scope}
\draw[line width=0.15mm,->] (-1,-3) to (-1,3) ;
\draw[line width=0.15mm,->] (-1,2) to (7.5,2) ;
\node at (-1,3.5) {$\mathrm{Im}[t]$}; 
\node at (8.5,2) {$\mathrm{Re}[t]$};
\draw[line width=0.35mm] (-1,2) to (6,2) to [out=0, in =0] (6,1) to (-1,1) to (-1,0) to (5,0) to [out=0, in =0] (5,-1) to (-1,-1) to (-1,-2);
\draw (6.1,2) node[cross=4pt,line width=0.35mm,red] {};
\draw (6.1,1.9) node[above] {$\scriptstyle W(t_1)$};

\node at (-2.,1) {$\scriptstyle-i {\beta\over 4}$}; 
\draw[dotted,line width=0.25mm] (-1.45,1) to (-1.1,1) ;
\draw (-1,1) node[cross=4pt,line width=0.35mm,blue] {};
\node at (-0.45,0.67) {$\scriptstyle V$};

\node at (-2,-0) {$\scriptstyle-i {\beta\over 2}$}; 
\draw[dotted,line width=0.25mm] (-1.45,0) to (-1.1,0) ;
\draw (5.1,0) node[cross=4pt,line width=0.35mm,red] {};
\draw (5.1,-0.1) node[above] {$ \scriptstyle W(t_2)$};

\node at (-2,-1) {$\scriptstyle-i {3\beta\over 4}$}; 
\draw[dotted,line width=0.25mm] (-1.45,-1) to (-1.1,-1) ;
\draw (-1,-1) node[cross=4pt,line width=0.35mm,blue] {};
\node at (-0.45,-1.3) {$\scriptstyle V$};

\node at (-2,-2) {$\scriptstyle-i\beta$}; 
\draw[dotted,line width=0.25mm] (-1.45,-2) to (-1.1,-2) ;
\draw[line width=0.35mm, ->] (-1,2) to (1.5,2) ;
\draw[line width=0.35mm, ->] (5,1) to (2.5,1) ;
\draw[line width=0.35mm, ->] (-1,0) to (1.5,0) ;
\draw[line width=0.35mm, ->] (5,-1) to (2.5,-1) ;
\draw[line width=0.35mm, ->] (-1,1) to (-1,0) ;
\draw[line width=0.35mm, ->] (-1,-1) to (-1,-2) ;

\node at (3,-3.) {\bf (a)}; 
\end{scope}
\begin{scope}
[xshift = 12.5cm]
\draw[line width=0.15mm,->] (-1,-3) to (-1,3) ;
\draw[line width=0.15mm,->] (-1,2) to (7.5,2) ;
\node at (-1,3.5) {$\mathrm{Im}[t]$}; 
\node at (8.5,2) {$\mathrm{Re}[t]$};
\draw[line width=0.35mm] (-1,2) to (6,2) to [out=0, in =0] (6,1-0.7) to (-1,1-0.7) to (-1,0-.5) to (5,0-.5) to [out=0, in =0] (5,-1-.2) to (-1,-1-.2) to (-1,-2);
\draw (6.1,2) node[cross=4pt,line width=0.35mm,red] {};
\draw (6.1,1.9) node[above] {$\scriptstyle W(t_1)$};

\begin{scope}
[yshift = -.7cm]
\node at (-2.6,1) {$\scriptstyle-i (1-\alpha-\sigma)\beta$}; 
\draw (-1,1) node[cross=4pt,line width=0.35mm,blue] {};
\node at (-0.45,0.67) {$\scriptstyle V$};
\end{scope}

\begin{scope}
[yshift = -.5cm]
\node at (-2.5,-0) {$\scriptstyle-i (1-\alpha)\beta$}; 
\draw[dotted,line width=0.25mm] (-1.25,0) to (-1.1,0) ;
\draw (5.1,0) node[cross=4pt,line width=0.35mm,red] {};
\draw (5.1,-0.1) node[above] {$ \scriptstyle W(t_2)$};
\end{scope}

\begin{scope}
[yshift = -.2cm]
\node at (-2.5,-1) {$\scriptstyle-i (1-\sigma)\beta$}; 
\draw[dotted,line width=0.25mm] (-1.25,-1) to (-1.1,-1) ;
\draw (-1,-1) node[cross=4pt,line width=0.35mm,blue] {};
\node at (-0.45,-1.3) {$\scriptstyle V$};
\end{scope}

\node at (-2.5,-2) {$\scriptstyle-i\beta$}; 
\draw[dotted,line width=0.25mm] (-1.85,-2) to (-1.1,-2) ;
\draw[line width=0.35mm, ->] (-1,2) to (1.5,2) ;
\draw[line width=0.35mm, ->] (5,1-.7) to (2.5,1-.7) ;
\draw[line width=0.35mm, ->] (-1,0-.5) to (1.5,0-.5) ;
\draw[line width=0.35mm, ->] (5,-1-.2) to (2.5,-1-.2) ;

\node at (3,-3.) {\bf (b)}; 
\end{scope}
\end{tikzpicture}
\vspace{-5mm}
\caption{(a) Extended Schwinger-Keldysh contour  corresponding to
  $\Tr[\rhoq V\rhoq W(t_2)\rhoq V \rhoq W(t_1) ]$. (b) Contour
  corresponding to a general regularization of the OTOC
  $\Tr[\rho^{\sigma}V\rho^{\alpha-\sigma} W(t_2)\rho^{\sigma} V
  \rho^{1-\alpha-\sigma} W(t_1)]$, which contributes to $C_{(\alpha,\sigma)}(t_1,t_2)$.}\label{fig:SK_sigma}
\end{figure}
\end{center}

We also point out that the dependence of the prefactor on the contour seems to be in tension with the recent attempts to associate \textit{maximal chaos}, defined as maximal Lyapunov exponent  $\lambda = 2\pi /\beta$, to \textit{destructive interference} of the commutator-squared \cite{Kitaev2017,Blake2017}. 
The {destructive interference} refers to the fact
that, if the decoherence factor equals $\cos(\lambda \beta/4)$,
it vanishes for
maximal chaos $\lambda=2\pi/\beta$. This implies  that for
maximal chaos the exponential time-dependence should be absent in the
symmetric commutator-squared. Our derivation shows, however, that this heavily relies on the analytical continuation chosen. Indeed, the
decoherence factor of commutator-squared of  Eq. \eqref{eq:Csigma} is
$\cos((1/4-\sigma)\lambda\beta)$, which does not vanish for maximal
chaos $\lambda=2\pi/\beta$, provided $0<\sigma\leq 1/4$.  This casts
doubts on how universal the relation between maximal chaos and destructive
interference may be.
  
Moreover, it has also been suggested that in SYK the prefactor of the
OTOCs $A \cos(\lambda\beta/4)$, where $A=\beta J/N$, is an observable
which is finite at zero temperature \cite{Gu2018}. However, as we have
shown above this quantity is contour-dependent and therefore, it is not
 an obvious physical observable.

\bigskip

To summarize, since the commutator-squared depends on the contour it is
not clear whether the regularised commutator-squared is actually an
observable. Another possibility may be that not all regularizations of
the commutator-squared are physically allowed and one value of $\sigma$
is preferred. For the specific deformation parametrized by $\sigma$, 
we could not find an argument for such case.

\subsection{The $\alpha$-contour}\label{sec:alpha_family}

Starting from $\alpha=1/2$, 
the parameter $\sigma$ affects only the decoherence factor of the
commutator-squared but leaves the Lyapunov spectrum invariant. There is therefore
a possibility that the Lyapunov spectrum as defined through the OTOC does measure a physical quantity. We set
$\sigma=0$ from here on and now explore
its dependence on the other contour parameter $\alpha$ which fixes the distance between the forwards branches, as shown in Fig  \ref{fig:SK_alpha}:
\begin{align}
&\Calpha=
\Tr\left[{\CA^\dagger\CA}
\right]\geq0\,,\quad  \CA\equiv  \rho^{1-\alpha\over2}\ [W(t),V(0)] \  \rho^{{\alpha\over2}}\,.\label{eq:Ct_alpha}
\end{align}

We have already seen that, for $\alpha\neq \frac{1}{2}$,  different
choices of $\sigma$ cannot be related by analytic
continuation. Neither can $\Calpha$ and $C(t;\beta)_{(\alpha',0)}$ be related to each other by analytic continuation. In other words, the distance in imaginary time between the forwards branches cannot be compensated by analytic continuation of time. This may be seen  explicitly by rewriting the OTOCs in $\Calpha$ as follows
\begin{align}
&\mathscr{H}_\alpha(t_1,t_2;t_3,t_4)\equiv 
\Tr\left[
\rhoS{\alpha}  V(t_3)W(t_1) \rhoS{1-\alpha} V(t_4)W(t_2)
\right]\nonumber\\
&=\Tr\left[
 V\left(t_3+i\beta\Big(\alpha{-}\quarter\Big)\right)\rhoq 
 W\left(t_1+i\beta\Big(\alpha{-}\half\Big)\right)\rhoq 
 V\left(t_4+i{\beta\over 4}\right)\rhoq 
 W(t_2)\rhoq 
\right]\,,\notag\\
&\mathscr{G}_\alpha(t_1,t_2)\equiv
\Tr\left[
\rhoS{\alpha} W(t_1) V(t_3) \rhoS{1-\alpha} W(t_2)V(t_4)
\right]\nonumber\\
&=\Tr\left[
 W\left(t_1+i\beta\Big(\alpha{-}\quarter\Big)\right)\rhoq 
 V\left(t_3+i\beta\Big(\alpha{-}\half\Big)\right)\rhoq 
 W\left(t_2+i{\beta\over 4}\right)\rhoq 
 V(t_4)\rhoq 
\right],\label{eq:weird_OTOC}
\end{align}
where we have chosen to compare to the standard contour with
$\rho^{1/4}$ separation. The differences between the complexified
times, 
$t_1+i\beta(\alpha-\frac{1}{2}),t_2,t_3+i\beta(\alpha-\frac{1}{4})$ and $t_4+i\beta/4$ in Eq. \eqref{eq:weird_OTOC}, no longer vanish in the analytically continued
OTOCs and this prevents relating one Lyapunov exponent to another.
In particular, the imaginary-time  separation between the two
operators $V(0)$ in both $\mathscr{G}_\alpha$  and
$\mathscr{H}_\alpha$  depends on $\alpha$. The standard choice,
$F(t)_{\left(\half,\quarter\right)}$, which is the building block used
to derive the bound on the Lyapunov exponent \cite{Maldacena2016c}, is
computed on a contour where the separation is $\beta/2$ and
$\alpha=1/2$. Therefore, $\mathscr{G}_\alpha$  and
$\mathscr{H}_\alpha$ cannot be related to
$F(t)_{\left(\half,\quarter\right)}$ by a simple analytic continuation
whenever $\alpha\neq1/2$ and we have to study the behavior of these
OTOCs separately. 

\begin{center}
 \begin{figure}[t!]
 \centering
\begin{tikzpicture}[scale=0.7][h]
\begin{scope}

\draw[line width=0.15mm,->] (-1,-3) to (-1,3) ;
\node at (-1,3.5) {$\mathrm{Im}[t]$}; 
\node at (9,2) {$\mathrm{Re}[t]$};
\draw[line width=0.15mm,->] (-1,2) to (8,2) ;

\draw[line width=0.35mm] (-1,2) to (6,2) to [out=0, in =0] (6,1.8) to (-1,1.8) to (-1,0.5) to (5,0.5) to [out=0, in =0] (5,0.3) to (-1,0.3) to (-1,-2);
\draw (6.05,2) node[cross=4pt,line width=0.35mm,red] {};
\draw (6.1,1.9) node[above] {$\scriptstyle W(t_1)$};

\draw (-1,1.8) node[cross=4pt,line width=0.35mm,blue] {};
\node at (-0.5,1.4) {$\scriptstyle V$};

\draw (5.05,0.5) node[cross=4pt,line width=0.35mm,red] {};
\draw (5.1,0.5) node[above] {$ \scriptstyle W(t_2)$};

\draw (-1,0.3) node[cross=4pt,line width=0.35mm,blue] {};
\node at (-0.5,-0.3) {$\scriptstyle V$};

\draw[line width=0.35mm, ->] (-1,2) to (1.5,2) ;
\draw[line width=0.35mm, ->] (5,1.8) to (2.5,1.8) ;
\draw[line width=0.35mm, ->] (-1,0.5) to (1.5,0.5) ;
\draw[line width=0.35mm, ->] (3,0.3) to (2.5,0.3) ;
\draw[line width=0.35mm, ->] (-1,1) to (-1,0.9) ;
\draw[line width=0.35mm, ->] (-1,-1) to (-1,-1.2) ;

\node at (-2.5,0.5) {$-i {\beta\alpha}$}; 
\draw[dotted,line width=0.25mm] (-1.75,0.5) to (-1.1,0.5) ;

\node at (-2.5,-2) {$-i {\beta}$}; 
\draw[dotted,line width=0.25mm] (-1.85,-2) to (-1.1,-2) ;

\end{scope}
\end{tikzpicture}
\vspace{-4mm}
	\caption{Extended Schwinger-Keldysh contour  corresponding to $tr[\rhoS{\alpha} W(t_1)V\rhoS{1-\alpha}W(t_2) V ]$ which enters in $\Calpha$ defined in Eq. \eqref{eq:Ct_alpha}. } \label{fig:SK_alpha}
\end{figure}
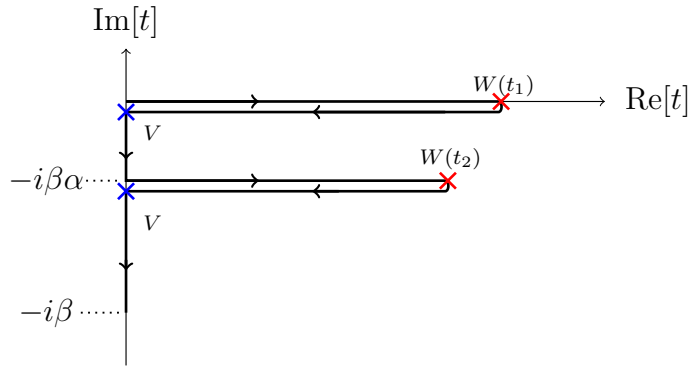
\end{center}

\subsection{OTOCS and physical observables in SK formalism}

As one may extrapolate from the previous section, the OTOC and its
Lyapunov spectrum will in general depend on the Schwinger-Keldysh contour on which it is computed. 
At first, this result may be surprising because, in standard Schwinger-Keldysh, it is known that \textit{physical} Green's functions  are independent of the contour due to energy conservation \cite{Matsumoto1983,Matsumoto1984}.   
Indeed, since the doubling of the contour is an artificial
mathematical convenience,  {a priori}  only correlation functions with external insertions on a single branch should considered physical, e.g.,
\begin{equation}
\langle \CO^{(1)}(x_1)\CO^{( 1)}(x_2)\CO^{(1)}(x_3)\rangle,\quad \langle \CO^{(2)}(x_1)\CO^{(2)}(x_2)\CO^{(2)}(x_3)\CO^{(2)}(x_4)\rangle\,,
\end{equation}
where we indicated with $(i)$ the branch where each operator is
inserted. With this definition, the fact that the correlation
functions do not depend on the contour is a simple diagrammatic proof.
We restate it here for the sake of clarity; it can be found in
\cite{Matsumoto1983,Matsumoto1984}. 

By inspecting the SK effective action, we know that the interaction vertices are of the form 
\begin{equation}
\mathcal{L}_{int}=\mathcal{L}_{int}^{(1)}-\mathcal{L}_{int}^{(2)}.
\end{equation}
Consequently, in the diagrammatic expansion each vertex is either of
type $1$ or of type $2$. The external legs of the vertices are
connected to each other or to external operator insertions with the propagators 
\begin{equation}
\langle\phi^{(i)}(-k)\phi^{(j)}(k)\rangle = \left(\begin{matrix}
    G_{\text{Feynman}} (k) & G^<_{\text{Wightman}}(k) \\
G^>_{\text{Wightman}} (k) & G_{\text{anti-Feyman}}(k)
\end{matrix}\right)_{ij}
\end{equation} 
Without loss of generality, we focus on the simple lowest order 1PI
diagram with $n$ operators inserted the branch $1$ and only one
$n$-point vertex:  $\langle \CO_1^{(1)}(k_1)....\CO_n^{(1)}(k_n)
\rangle_\alpha$. 
Clearly if
the vertex is of type 1, there is no contour dependence in the
diagram. When the vertex is of type 2, as in Fig. \ref{fig:OTOC_sketch}, we need to use a Wightman
function. For a general contour where the forward and backward
branches are separated by $\rho^{\alpha}$ this is one of the Wightman
functions\footnote{In the literature the following notation is sometimes used: $G_{12}(k)=G^{<}(k)$ and $G_{21}(k)=G^>(k)$.}
\begin{align}
  \label{eq:3}
  G^{\beta\alpha}_{12}(k) &= \text{Tr} \rho^{1-\alpha}
  \phi^{(1)}(-k)\rho^{\alpha}\phi^{(2)}(k)  \nonumber\\
G^{\beta\alpha}_{21}(k) &= \text{Tr} \rho^{1-\alpha}
  \phi^{(2)}(-k)\rho^{\alpha}\phi^{(1)}(k)  = e^{\beta(1-2\alpha) k^0}G^{\beta\alpha}_{12}(k)
\end{align}
By Fourier transforming the time direction, using
$\rho^{\alpha}\hat{O}(t)\rho^{-\alpha} = \hat{O}(t+i\alpha)$ and
Fourier transforming back, one readily derives that
\begin{align}\label{eq:widthDep}
G_{12}^{\beta\alpha}(k) &= e^{\beta\alpha
  k^0}G_{12}^{\alpha=0}(k), \nonumber\\
G_{21}^{\beta\alpha}(k) &= e^{-\beta\alpha k^0}G_{21}^{\alpha=0}(k).
\end{align} 
At lowest order, there is a
single $n$-point vertex on branch 2. Contracting each of the legs of
the vertex with the external operators on branch 1, and by using \eqref{eq:widthDep},
this means that the relation between correlation function on different
contours is
\begin{equation}
\langle \CO_1^{(1)}(k_1)....\CO_n^{(1)}(k_n) \rangle_\alpha \sim e^{\beta\alpha\sum_{i=1,..,n} k^{0}_i } \langle \CO_1^{(1)}(k_1)....\CO_n^{(1)}(k_n)  \rangle_{\alpha=0}=\langle \CO_1^{(1)}(k_1)....\CO_n^{(1)}(k_n)  \rangle_{\alpha=0}
\end{equation}
Because of energy conservation at
the vertex, $\sum_{i=1,..,n} k_i = 0$, the overall factor vanishes and
this proves the contour independence of these types of  diagrams.

\begin{figure}[!t]
\begin{center}
\begin{tikzpicture}[scale=0.8][h]
\begin{scope}
\node at (-0.5,1.5) {$\CO_1^{(1)}$};
\draw[line width=0.35mm] (-0.3,1.2) to (2,0) ;
\node at (0.8,1.1) {$\mathbf{k_1}$}; 

\node at (4.9,1.5) {$\CO_2^{(1)}$};
\draw[line width=0.35mm] (2,0) to (4.3,1.2) ;
\node at (3.2,1.1) {$\mathbf{k_2}$}; 

\node at (-0.5,-1.5) {$\CO_n^{(1)}$};
\draw[line width=0.35mm] (-0.3,-1.2) to (2,0) ;
\node at (0.8,-1.1) {$\mathbf{k_n}$}; 

\node at (4.9,-1.5) {$\CO_3^{(1)}$};
\draw[line width=0.35mm] (2,0) to (4.3,-1.2) ;
\node at (3.2,-1.1) {$\mathbf{k_3}$}; 

\shade[left color=Cyan](2,0) circle (15pt);
\node at (2,0) {$V_{2}$}; 

\draw[dotted,dash pattern=on \pgflinewidth off 5pt,line width=0.4mm] (1.5,-1.2) to (2.5,-1.2);


\end{scope}

\begin{scope}[xshift=10cm, scale=0.8]

\draw[line width=0.15mm,->] (-1,-3) to (-1,3) ;
\node at (-1,3.5) {$\mathrm{Im}[t]$}; 
\node at (9,2) {$\mathrm{Re}[t]$};
\draw[line width=0.15mm,->] (-1,2) to (8,2) ;

\draw[line width=0.35mm] (-1,2) to (7,2) to  (7,0.3)  to (-1,0.3) to (-1,-2);
\draw (6.05,2) node[cross=4pt,line width=0.35mm,blue] {};
\draw (6.1,2.1) node[above] {$\scriptstyle \CO_3$};

\draw (4.05,2) node[cross=4pt,line width=0.35mm,blue] {};
\draw (4.1,2.1) node[above] {$\scriptstyle \CO_1$};
\draw (0.05,2) node[cross=4pt,line width=0.35mm,blue] {};
\draw (0.1,2.1) node[above] {$\scriptstyle \CO_2$};
\draw (2.05,2) node[cross=4pt,line width=0.35mm,blue] {};
\draw (2.1,2.1) node[above] {$\scriptstyle \CO_n$};

\shade[left color=Cyan](3.5,0.3) circle (15pt);
\node at (3.5,0.3) {$V_{2}$};

\draw[line width=0.35mm, ->] (-1,2) to (1.5,2) ;
\draw[line width=0.35mm, ->] (3,0.3) to (2.5,0.3) ;
\draw[line width=0.35mm, ->] (7,1) to (7,0.9) ;
\draw[line width=0.35mm, ->] (-1,-1) to (-1,-1.2) ;

\node at (-2.5,0.3) {$-i {\beta\alpha}$}; 
\draw[dotted,line width=0.25mm] (-1.75,0.3) to (-1.1,0.3) ;

\node at (-2.5,-2) {$-i {\beta}$}; 
\draw[dotted,line width=0.25mm] (-1.85,-2) to (-1.1,-2) ;

\end{scope}
\end{tikzpicture}

\caption{A diagrammatic expansion of the correlator with the external legs on the same branch of the SK contour. The result does not depend on the width $\beta\alpha$.}\label{fig:OTOC_sketch}
\end{center}
\end{figure}
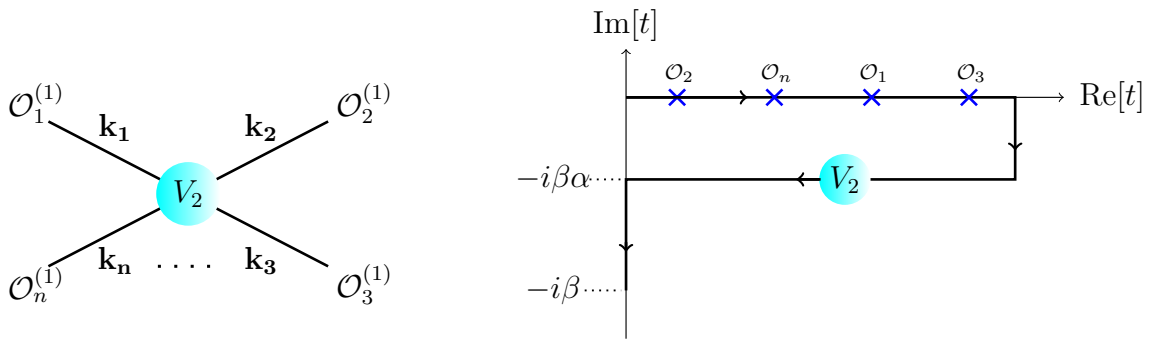
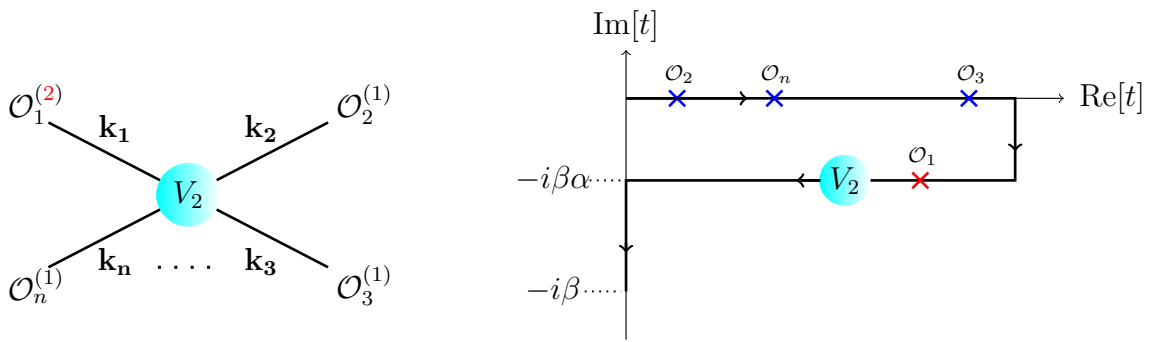
\begin{figure}[!t]
\begin{center}
\begin{tikzpicture}[scale=0.8][h]

\begin{scope}
\node at (-0.5,1.5) {$\CO_1^{(\color{red}2\color{black})}$};
\draw[line width=0.35mm] (-0.3,1.2) to (2,0) ;
\node at (0.8,1.1) {$\mathbf{k_1}$}; 

\node at (4.9,1.5) {$\CO_2^{(1)}$};
\draw[line width=0.35mm] (2,0) to (4.3,1.2) ;
\node at (3.2,1.1) {$\mathbf{k_2}$}; 

\node at (-0.5,-1.5) {$\CO_n^{(1)}$};
\draw[line width=0.35mm] (-0.3,-1.2) to (2,0) ;
\node at (0.8,-1.1) {$\mathbf{k_n}$}; 

\node at (4.9,-1.5) {$\CO_3^{(1)}$};
\draw[line width=0.35mm] (2,0) to (4.3,-1.2) ;
\node at (3.2,-1.1) {$\mathbf{k_3}$}; 

\shade[left color=Cyan](2,0) circle (15pt);
\draw[dotted,dash pattern=on \pgflinewidth off 5pt,line width=0.4mm] (1.5,-1.2) to (2.5,-1.2);

\node at (2,0) {$V_{2}$}; 

\end{scope}
\begin{scope}[xshift=10cm, scale=0.8]

\draw[line width=0.15mm,->] (-1,-3) to (-1,3) ;
\node at (-1,3.5) {$\mathrm{Im}[t]$}; 
\node at (9,2) {$\mathrm{Re}[t]$};
\draw[line width=0.15mm,->] (-1,2) to (8,2) ;

\draw[line width=0.35mm] (-1,2) to (7,2) to  (7,0.3)  to (-1,0.3) to (-1,-2);
\draw (6.05,2) node[cross=4pt,line width=0.35mm,blue] {};
\draw (6.1,2.1) node[above] {$\scriptstyle \CO_3$};

\draw (5.05,0.3) node[cross=4pt,line width=0.35mm,red] {};
\draw (5.1,0.4) node[above] {$\scriptstyle \CO_1$};
\draw (0.05,2) node[cross=4pt,line width=0.35mm,blue] {};
\draw (0.1,2.1) node[above] {$\scriptstyle \CO_2$};
\draw (2.05,2) node[cross=4pt,line width=0.35mm,blue] {};
\draw (2.1,2.1) node[above] {$\scriptstyle \CO_n$};

\shade[left color=Cyan](3.5,0.3) circle (15pt);
\node at (3.5,0.3) {$V_{2}$};

\draw[line width=0.35mm, ->] (-1,2) to (1.5,2) ;
\draw[line width=0.35mm, ->] (3,0.3) to (2.5,0.3) ;
\draw[line width=0.35mm, ->] (7,1) to (7,0.9) ;
\draw[line width=0.35mm, ->] (-1,-1) to (-1,-1.2) ;

\node at (-2.5,0.3) {$-i {\beta\alpha}$}; 
\draw[dotted,line width=0.25mm] (-1.75,0.3) to (-1.1,0.3) ;

\node at (-2.5,-2) {$-i {\beta}$}; 
\draw[dotted,line width=0.25mm] (-1.85,-2) to (-1.1,-2) ;

\end{scope}

\end{tikzpicture}

\caption{A diagrammatic expansion of the correlator with all but one  external legs on the same branch of the SK contour. The result does depend on the width $\beta\alpha$ }\label{fig:OTOC_sketch2}
\end{center}
\end{figure}

However, if one of the external legs is in the branch 2, see
Fig. \ref{fig:OTOC_sketch2}, it is easy to
see that now one of the Green's function no longer depends on the
separation $\alpha$ at all, so the global factor in
the $n$ point function does not simplify anymore. The simplest example
of this is the Wightman function itself. There is no vertex, but we
have already shown that $G_{12}^{\alpha} \neq G_{12}^{0}$ above in Eq.~\eqref{eq:3}.
Extending to an $n$-point correlation functions with a single $n$-point vertex,
one has
\begin{equation}
\langle \CO_1^{(\color{red}2\color{black})}(k_1)....\CO^{(1)}_n(k_n) \rangle_\alpha \sim e^{\beta\alpha\sum_{i=2,..,n} k^{0}_i } \langle \CO_1^{(\color{red}2\color{black})}(k_1)....\CO^{(1)}_n(k_n)  \rangle_{\alpha=0}\neq\langle \CO_1^{(\color{red}2\color{black})}(k_1)....\CO^{(1)}n(k_n)  \rangle_{\alpha=0}~,
\end{equation}
but now the exponent in the prefactor $\sum_{i=2,..,n} k^{0}_i  =
-k^0_1 \neq 0$.

It is not difficult to see that the simple proof shown above extends to any diagrams. Indeed, given any diagram of the expansion, it is sufficient to divide it in subdiagrams and to use the momentum conservation in each vertex.

Turning our attention back to the OTOC, by construction each insertion
occurs on one of four different branches. This indicates that the OTOC
will be contour dependent, similar to two-branch correlation function
in Schwinger-Keldysh theory as depicted in Fig.~\ref{fig:OTOC_sketch2}.
If so, this does not immediately mean that the OTOC does not measure a
physical quantity (in part). For example,  the (bosonic) Wightman function $G_{12}^{\beta\alpha}(k) = e^{\beta k^0}(1+n(k^0))\rho(k)$ depends on the contour, but still encodes a physical quantity, namely the spectral density $\rho(k)$. Therefore, more care is needed to understand the
relation between the contour-dependent OTOC and physical properties of the system.

\section{Contour dependence of the Lyapunov spectrum in a matrix
  $\Phi^4$ theory at weak coupling}\label{sec:weak}

We now prove by direct computation that the OTOC indeed depends on the 
details on the contour chosed. In this section, we compute Lyapunov
spectrum obtained from the  commutator-squared $\Calpha$ in a perturbative matrix
field theory, which has been studied in detail for $\alpha=1/2$ in
\cite{Stanford2016}. The advantage of the perturbative field theory calculation
is that the commutator-squared can be related to a kinetic equation encoding the
microscopic dynamics \cite{Grozdanov:2018atb}. From this, we will suggests that this microscopic
insight argues that one specific contour, the one with $\alpha=1/2$ is
the one that computes microscopic chaos.

\noindent We consider a 3+1 dimensional QFT with a  Hermitian matrix field
$\Phi_{ab}$ whose Lagrangian is  given by 
\begin{equation}
\mathcal{L} = 
\Tr\left( \half\dot{\Phi}^2 -\half(\nabla\Phi)^2 - \half m^2 \Phi^2 - {g^2\over 4!} \Phi^4\right)\,,
\end{equation}
with $g^2 = \lambda N$.

\noindent The commutator-squared of Eq. \eqref{eq:Ct_alpha} in this matrix model is
\begin{equation}
\Calpha = \frac{1}{N^4}\sum_{aba'b'}\int d^3\mathbf{x} \ \Tr\Big( \rhoS{1-\alpha}\, [\Phi_{ab}(t,\mathbf{x}),\Phi_{a'b'}]\rhoS{\alpha}\, [\Phi_{ab}(t,\mathbf{x}),\Phi_{a'b'}]^\dag\Big).
\end{equation}
For $t>0$, which we shall assume, the lowest order (disconnected)
contribution is the product of two retarded Green's function arising
from a contraction on the top two folds and the bottom folds
separately; there is therefore no contour dependence. The
non-trivial contribution at the next order, that can seed exponential
growth, is the contribution with two Wightman functions connecting the
two retarded Green's functions.
For $\alpha=1/2$, this equals \cite{Stanford2016}:
\begin{eqnarray}
C(\omega)_{(\half,0)}^{(1)} = \frac{1}{N^2}\int 
{\dd^4p\over (2\pi)^4}{\dd^4p'\over (2\pi)^4} G_R(\omega-p)G_R(p)R(p-p')G_R(\omega-p')G_R(p')\,,\label{eq:BS_Stanford}
\end{eqnarray}
where the kernel $R(p)$ is determined in terms of Wightman functions with operators separated by $i\beta/2$:
\begin{eqnarray}
R(p) = \frac{g^4(N^2+5)}{12}\int \frac{d^4\ell}{(2\pi)^4}G_{12}^{\beta/2}(p/2 + \ell)G_{12}^{\beta/2}(p/2 - \ell)\,,\label{eq:kernel_Matrix}
\end{eqnarray}
Note that it is only $G_{12}(k)$ and not $G_{21}(k)$, independent of
the deformation $\alpha=1/2$, which appears inside the kernel. This
choice is due to the identity $G_{12}(k)=G_{21}(-k)$. We will
consistently use $G_{12}$ only; this will not affect the final result.
Defining a function $f(\omega, p)$,
\begin{equation}\label{eq:f_matrix}
C(\omega)_{(\half,0)}= \frac{1}{N^2}\int \frac{d^4p}{(2\pi)^4}f(\omega,p)~,
\end{equation}
at the next order one of the contributions is
\begin{eqnarray}
C(\omega)_{(\half,0)}^{(2)} = \frac{1}{N^2}\int 
{\dd^4p\over (2\pi)^4}{\dd^4p'\over (2\pi)^4} G_R(\omega-p)G_R(p)R(p-p')f^{(1)}(\omega,p')\,,\label{eq:BS_Stanford_2}
\end{eqnarray}
and by rewriting $C(\omega)_{(\half,0)}^{(2)} =\frac{1}{N^2}\int \frac{d^4p}{(2\pi)^4}f^{(2)}(\omega,p)$, one can set up a recursive Bethe-Salpeter equation to determine
$f(\omega,p)$ and hence $C(\omega)_{(\half,0)}$ to all orders. Since
we are interested in the late-time exponential growth, we focus on the
homogeneous part of the Bethe-Salpeter equation, which in the 
low-frequency, late time limit equals
\begin{align}
f(\omega,p)&\simeq -G^R(p)G^R(\omega-p) 
\int {\dd^4 k\over (2\pi)^4} R(k-p)f(\omega,k)\,,\label{eq:BS_matrix1}
\end{align}
Equation~\eqref{eq:BS_matrix1} is only appropriate in the low $\omega$ limit. In
  this limit the product of retarded Green's functions is dominated by a
  {\it pinching pole singularity}, which amounts to the following
  approximation 
\cite{Stanford2016}
\begin{align}
G^R(p)G^R(\omega-p)&={\pi\over E_\p}{\delta({p_0}^2-E_\p^2)\over
                     i\omega-2\Gamma_p}+ \ldots\,\,\,\,.
\end{align}
As this concentrates the support of the right hand side of the BSE on
the on-shell delta-function, there is natural ansatz for the solution
of $f(\omega,p)$ to be proportional to the same delta-function
\begin{align}
  \label{eq:18}
  f(\omega,p)_{\text{ansatz}} = f(\omega,\p)\delta(p_0^2-E_\p^2).
\end{align}
The imaginary part of the two-loop ($\alpha$-independent) 
self energy $\Gamma_p$ 
also happens to be determined in terms of (the $\alpha=1/2$) $R(k)$ defined in Eq. \eqref{eq:kernel_Matrix}:
\begin{equation}\label{eq:Gammap}
\Gamma_\p ={\sinh\left({\beta E_\p\over2}\right)\over 24 E_\p} \int{ \dd^3 k\over (2\pi)^3} 
{R(E_\p-E_\k,\p-\k)+R(E_\p+E_\k,\p-\k)\over E_\k \sinh\left({\beta E_\k\over2}\right)}\,.
\end{equation}
Including both the pinching pole approximation and the self-energy
rewriting in terms of the rung function $R(k-p)$, the low-energy approximation of the BSE reads
\begin{align}
  \label{eq:9}
 - i\omega f(\omega,\p) =  \int \frac{ \dd^3 k}{ (2\pi)^3}\left(\mathcal{K}(	\p,\k) -2 \Gamma_\k (2\pi)^3\delta^3(\p-\k)\right)f(\omega,\k),
\end{align}
where $\mathcal{K}(	\p,\k) = \frac{R(E_\p-E_\k,\p-\k) + R(E_\p+E_\k,\p-\k)}{4E_\k E_\p}$. The positive eigenvalues of the kernel 
  $\mathcal{K}(	\p,\k) -2 \Gamma_\p \delta^3(\p-\k) $, considered as
  a matrix 
in $k$ and $p$, form the Lyapunov spectrum characterizing the
exponential growth at late times, as we will review below. 

Importantly, the Lyapunov spectrum is {\it not} set by the off-shell rung
function $R(p-k)$ or the off-shell BSE
Eq.~\eqref{eq:BS_matrix} \cite{Grozdanov:2018atb}. Specifically, the
``on-shell'' delta-function ansatz for \mbox{$f(\omega,p) = f(\omega,\p)\delta(p_0^2-E_p)$}, that
naturally follows the pinching pole approximation, acts as a projector
on the set of functions $f(\omega,p)$. Therefore the set of
eigenvalues and eigenvectors of $R(k-p)$ are not the same as those of
the kernel in Eq.\eqref{eq:9} which sets the Lyapunov spectrum.  

\bigskip

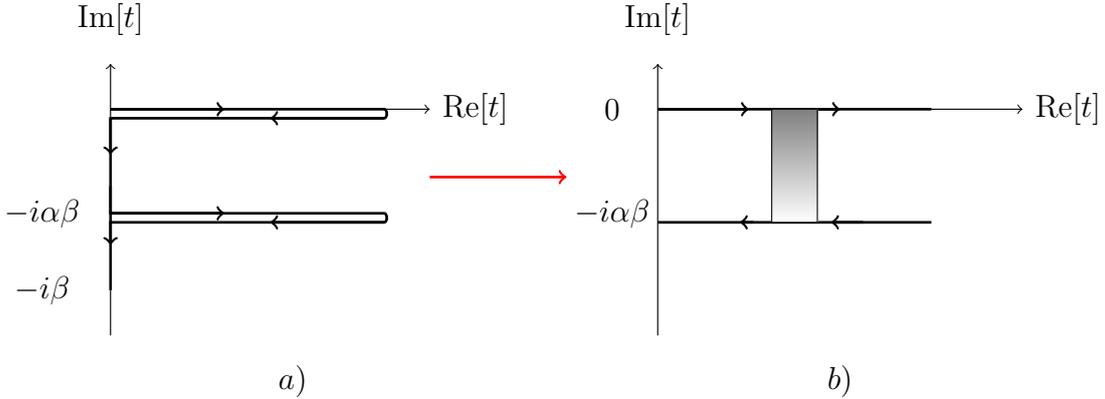
\begin{figure}[!h]
\begin{center}
\begin{tikzpicture}[scale=0.6][h]
\begin{scope}
\draw[line width=0.15mm,->] (-1,-3) to (-1,3) ;
\node at (-1,4) {$\mathrm{Im}[t]$}; 
\node at (7,2) {$\mathrm{Re}[t]$};
\draw[line width=0.15mm,->] (-1,2) to (6,2) ;
\draw[line width=0.35mm] (-1,2) to (5,2) to [out=0, in =0] (5,1.8) to (-1,1.8) to (-1,-0.3) to (5,-0.3) to [out=0, in =0] (5,-0.5) to (-1,-0.5) to (-1,-2);
\node at (-2.5,-0.3) {$-i\alpha \beta$}; 
\node at (-2.5,-2) {$-i\beta$}; 
\draw[line width=0.35mm, ->] (-1,2) to (1.5,2) ;
\draw[line width=0.35mm, ->] (5,1.8) to (2.5,1.8) ;
\draw[line width=0.35mm, ->] (-1,-0.3) to (1.5,-0.3) ;
\draw[line width=0.35mm, ->] (5,-0.5) to (2.5,-0.5) ;
\draw[line width=0.35mm, ->] (-1,1.8) to (-1,1) ;
\draw[line width=0.35mm, ->] (-1,0.3) to (-1,-1) ;
\node at (3,-4) {$a)$}; 
\end{scope}
\draw[line width=0.35mm,->, color=red] (6,0.5) to (9,0.5) ;

\begin{scope}
[xshift = 12cm]
\draw[line width=0.15mm,->] (-1,-3) to (-1,3) ;
\node at (-1,4) {$\mathrm{Im}[t]$}; 
\node at (8,2) {$\mathrm{Re}[t]$};
\draw[line width=0.15mm,->] (-1,2) to (7,2) ;
\draw[line width=0.35mm] (-1,2) to (5,2) ;
\draw[line width=0.35mm]  (-1,-0.5) to (5,-0.5);
\node at (-2.,-0.3) {$-i\alpha \beta$}; 
\node at (-2.,2) {$0$}; 
\draw[line width=0.35mm, ->] (-1,2) to (1.,2) ;
\draw[line width=0.35mm, ->] (2,2) to (3,2) ;
\draw[line width=0.35mm, ->] (1.1,-0.5) to (0.8,-0.5) ;
\draw[line width=0.35mm, ->] (3.5,-0.5) to (2.8,-0.5) ;
\node at (3,-4) {$b)$}; 
\shadedraw(1.5,2) rectangle (2.5,-0.5);
\end{scope}
\end{tikzpicture}
\caption{A pictorial representation of a general time contour ($a$) and of the 4-points function in the ladder approximation ($b$) . The external legs lay on the first time fold and the second time fold. On the contrary, the rung joins the two time folds and include Wightman functions which by definition are contour dependent. }\label{fig:rung}
\end{center}
\end{figure}

This derivation makes clear that the only $\alpha$-contour-deformation
dependence arises from the Wightman functions in the rung function.
It is then straightforward to derive the contour-dependence of the OTOC.
For $\alpha\neq1/2$, the rung function should be modified as sketched in Fig. \ref{fig:rung}.
Mathematically
\begin{eqnarray}
R(p) \to e^{\beta p^0 (\alpha-1/2)}R(p)\,.
\end{eqnarray}
Again, defining 
\begin{equation}\label{eq:f_matrix}
\CalphaOm = \frac{1}{N^2}\int \frac{d^4p}{(2\pi)^4}f(\omega,p),
\end{equation}
this will now obey the equation:
\begin{align}
f(\omega,p)&\simeq -G^R(p)G^R(\omega-p)
\int {\dd^4 k\over (2\pi)^4}
             e^{\eta(k^0-p^0)}R(k-p)f(\omega,k)\,,\label{eq:BS_matrix}
\end{align}
with $\eta\equiv\beta(\alpha-1/2)$. 
Note that the change in the rung function does not depend on 
  whether it is constructed from $G_{12}(k)$ or $G_{21}(-k)$. This can
  be confirmed by the fact that the commutator-squared should obey a 
KMS type symmetry $\alpha
  \rightarrow 1-\alpha$ on the
  doubled time contour. This follows by 
  redefining $k\rightarrow \omega -k$ and $p\rightarrow \omega-p$. The kernel $R(k-p)$
  is even in $k-p$ as can be readily seen from its definition
  Eq.~\eqref{eq:kernel_Matrix}. The 
  product $G^R(p)G^R(\omega-p)$ changes into itself, and one obtains
  an equation for $f(\omega;\omega-p^0,-\p)$ which identical to the
  original equation.

To solve the Bethe-Salpeter equation \eqref{eq:BS_matrix} after
  the pinching pole approximation in the late time limit,
\begin{align}
f(\omega,p)&\simeq {\pi\over E_\p}{\delta({p_0}^2-E_\p^2)\over
                     -i\omega+2\Gamma_p} 
\int {\dd^4 k\over (2\pi)^4}
             e^{\eta(k^0-p^0)}R(k-p)f(\omega,k)\,,\label{eq:BS_matrix2}
\end{align}
one then makes the natural ansatz
\begin{align}
f(\omega,p)&=f(\omega,\p) \delta(p_0^2-E_\p^2)\,.\label{eq:f_delta}
\end{align}
However, note that the choice of the ansatz is very subtle and might
lead to a different physical solution. By inspecting
eq. \eqref{eq:BS_matrix}, one might be tempted to argue that, since
the $\eta$ dependent term resembles a similarity transformation, the
eigenvalues are unchanged. This conclusion is not correct.  Indeed, as
briefly recalled above, we showed in \cite{Grozdanov:2018atb} that
there are physical implications in this choice. Most notably, the
$\eta = 0$ BSE with the kernel replacement $R(k-p)\to \frac{\mathrm{sinh}(\beta
  p_0/2))}{\mathrm{sinh}(\beta k_0/2)} R(k-p)$ corresponds to the
evaluation of a {\em different} analytic continuation of the
commutator-squared. This analytic continuation is the retarded
correlation function of the  Wigner transform of the bilocal density
operator, namely the correlation function that appears in the Kubo
formula of the shear viscosity. From Boltzmann's kinetic theory, the
spectrum with this ansatz is manifestly negative definite (\emph{i.e.} there are
only decaying modes). In other words, the choice of contour dictates
the ansatz to solve the BSE. At the same time, all the contours that
are related through analytic continuation arise from the same
contour-independent off shell BSE Eq.~\eqref{eq:BS_matrix2}.
 
Substituting this {\it appropriate} ansatz  \eqref{eq:f_delta} into
eq. \eqref{eq:BS_matrix}, we then perform the integral over
$p_0$. This yields
\begin{align}
&(-i\omega+2\Gamma_{p}) f(\omega,\p)= 
\int {\dd^3 k\over (2\pi)^3} f(\omega,\k) \mathcal{K}(\k,\p,\eta),\label{eq:mtilde}
\end{align}
with
\begin{align}
\mathcal{K}(\k,\p,\eta){\equiv}
{\cosh\left[ \eta(E_\p-E_\k)\right]R\big(E_\p-E_\k,\p-\k\big)
+\cosh\left[ \eta(E_\p+E_\k)\right] R\big(E_\p+E_\k,\p-\k\big)
\over 4E_\k E_\p }
\nonumber \,,
\end{align}
where we have explicitly used that the rung kernel is even in the energy argument: $R(k_0,\k)=R(-k_0,\k)$.
Substituting Eq. \eqref{eq:Gammap} into Eq. \eqref{eq:mtilde}, we
arrive at the final Bethe-Salpeter equation for  $C(\omega)_{(\alpha,0)}=\frac{1}{N^2} \int\frac{d^4p}{(2\pi)^4}f(\omega,p)$ in the frequency domain:
\begin{align}
-i\omega f(\omega,\p)&{=} 
\int \hspace{-2mm}{\dd^3 k\over (2\pi)^3} 
\left[f(\omega,\k) \mathcal{K}(\k,\p,\eta)
{-}f(\omega,\p)
{\sinh\left({\beta E_\p\over2}\right)\over \sinh\left({\beta E_\k\over2}\right)} 
{R\big(E_-,\p{-}\k\big ){+}R\big(E_+,\p{-}\k\big)\over 
12 E_\p E_\k }\right],\label{eq:BS_eta}
\end{align}
where $E_\pm \equiv E_\p \pm E_\k$.
In the time domain this is an equation of the type
\begin{align}
  \label{eq:5}
  \frac{\partial}{\partial t} f(t)_\p = {\cal M}_{\p\k}f_\k(t).
\end{align}
The solutions are the eigenvectors of ${\cal M}_{\p\k}$ with an
exponential growth/decay in time proportional to the eigenvalue. The
positive eigenvalues of ${\cal M}_{\p\k}$ are the Lyapunov
spectrum. This can be found numerically; the precise method used to
solve this equation may be found in Appendix \ref{{app:matrix}}. 
Without computation it is already clear, however, that the result will
depend on the $\alpha$-deformed contour, as the defining
Bethe-Salpeter equation does so.
\begin{figure}[t!]
\hspace{-15mm}
\includegraphics[scale=0.65]{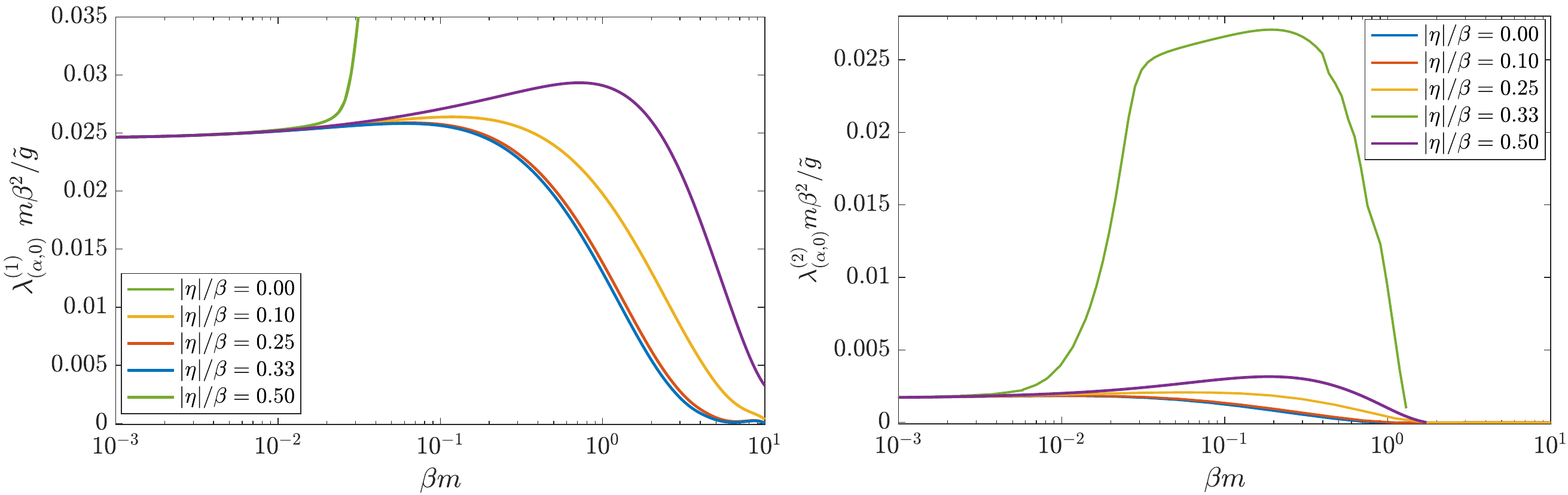}
\vspace{-9mm}
\caption{\textbf{Contour dependence of the Lyapunov spectrum in the weakly coupled $\Phi^4$-matrix model}. Two largest Lyapunov exponents of Lyapunov spectrum of the matrix $\Phi^4$ theory as a function of the coupling. Each value of $\eta$, defined in Eq. \eqref{eq:BS_matrix}, corresponds to a different contour choice: $\eta=0$ corresponds to the common symmetric regularization. {For brevity, we defined  $\tilde{g} =\frac{g^4(N^2+5)}{4\cdot 144} $.}}\label{fig:matrix}
\end{figure}

The result is presented in Fig. \ref{fig:matrix}. We clearly see the
dependence of the two positive Lyapunov exponents on the contour. The 
spectrum does become
contour-independent in the high-temperature limit. This follows
directly from the fact that the deformation parametrized by
$\eta=\beta(\alpha-1/2)$ becomes negligible for small $\beta$
(compared to the mass). 

That in these models the Lyapunov spectrum is contour independent
  for zero mass, will be crucial to understand the SYK model, which we
  study in the next section. There, there are only gapless excitations
  and not even a thermal mass,
  and we can therefore expect the same contour  independence of the
  Lyapunov spectrum as the
  $\beta m\to 0$ limit of weakly coupled field theories as exhibited
  in Fig.\ref{fig:matrix}. Do recall that the full OTOC always depends
on the contour.

\begin{figure}[t!]
\center
\includegraphics[scale=0.7]{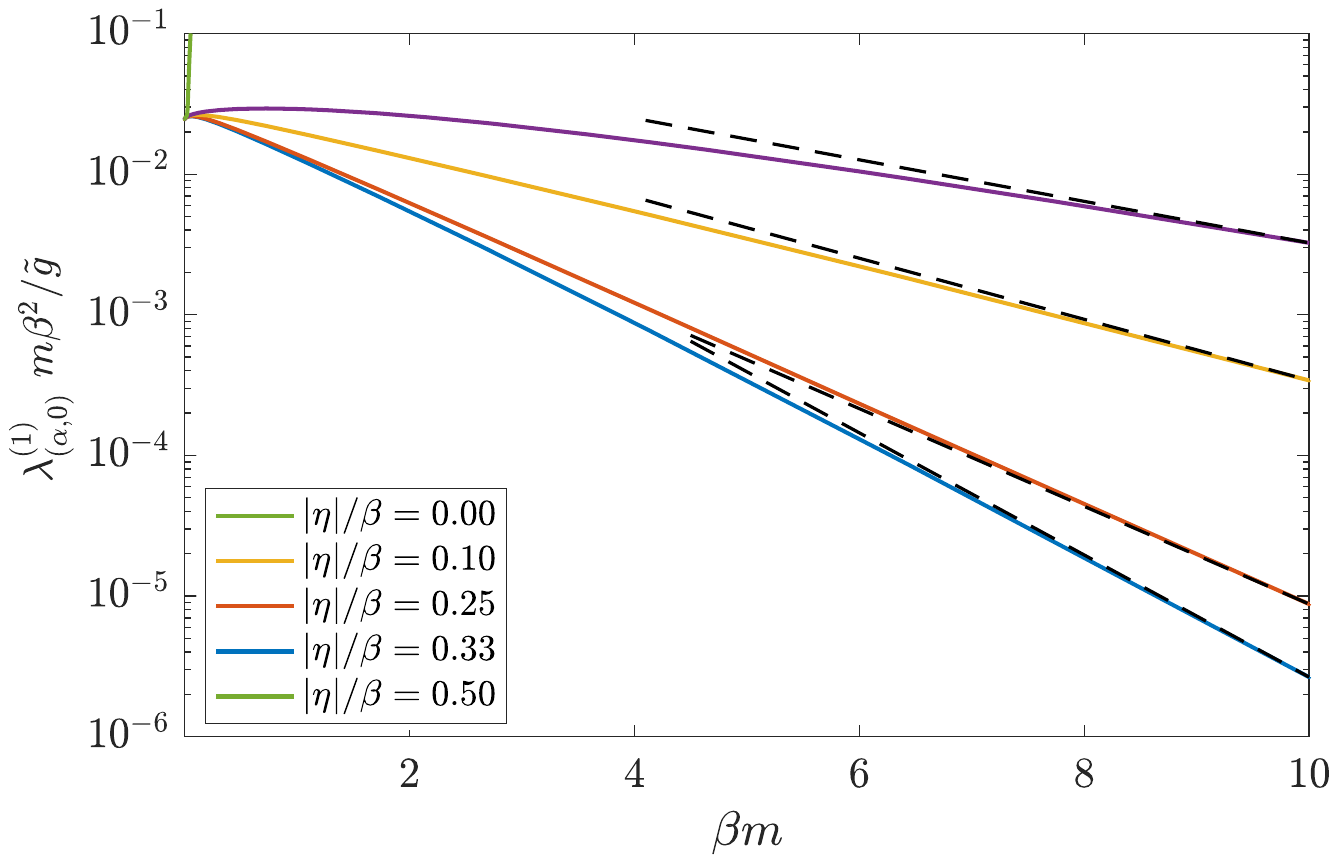}
\vspace{-3mm}
\caption{\textbf{Exponential decay of the first Lyapunov exponent for
    various contours}. Dashed lines correspond to the analytical
  expression $C e^{-(\beta +2|\eta|)m}$, where $C$ is fixed so that the dashed line passes through the last point available $\beta m=10$. For brevity, we defined  $\tilde{g} =\frac{g^4(N^2+5)}{4\cdot 144} $.}\label{fig:matrix2}
\end{figure}

For intermediate and small $\beta$,
the Lyapunov spectrum sensitively depends on the choice of
contour. As also noted already in \cite{Stanford2016}, in the extreme low
temperature limit $\beta m\to\infty$, the Lyapunov spectrum vanishes
exponentially in $\beta m$. Even though this decreases the relative
dependence on the contour, 
the contour dependence still persists and is given by
$e^{-(\beta+2|\eta|)m}$,
 as shown in Fig. \ref{fig:matrix2}.

Let us make one final comment on the connection between the
  choice of ansatz and the contour dependence of the Lyapunov
  spectrum. 
One readily observes
  that another possible ansatz to the BSE is 
\begin{align}
\label{eq:alt_ansatz}
f(\omega,p)&=f(\omega,\p) e^{\eta p^0}\delta(p_0^2-E_\p^2) \nonumber \\
&= f(\omega,\p)\left(e^{\eta E_\p}\frac{\delta(p^0 -E_\p)}{2E_\p}+e^{-\eta E_\p}\frac{\delta(p^0 +E_\p)}{2E_\p}\right).
\end{align}
This $\eta$-contour skewed ansatz gives a contour independent Lyapunov exponent as solution
for Eq.~\eqref{eq:BS_matrix2} and coincides with the solution for
$\eta=0$. One may ask why one ansatz is preferred over the other. As
is clear from  fig. \ref{fig:matrix}, the natural ansatz
\eqref{eq:f_delta} represents a solution with a larger eigenvalue of
the Lyapunov exponent. We therefore argue that this solution is what a general
computation, \emph{i.e.} using different techniques than the BSE,  of the leading exponential rate of growth in the OTOC
would capture. In support of this, we also refer to the results of
\cite{Liao:2018uxa}. There, the authors computed the OTOC for a $2+1$
disordered systems by means of a Keldysh nonlinear sigma model
technique that they developed. Within this framework, the computation of
the largest exponent for the unregularised $\eta=1/2$ case and the
regularised $\eta=0$ case gives a different result. Moreover, the
unregularised case has a larger exponetial growth rate. The
explanation is the one we give above.

This is the message to take from these results. When one computes the OTOC,
one is inherently concerned with the late time regime of the
correlator and with the largest term in the exponential
growth. Mathematically the Lyapunov exponent of the fastest growing mode is contour
dependent. This gives rise to the physics question of how we can understand the different contour dependent growing rates. We will answer this question in sec. \ref{sec:kinet-theory-interpr}.

\subsection{The contour dependence regulates the IR}

The contour dependence of the Lyapunov spectrum explicitly exhibited
above emphasizes an important point regarding the physics behind the
contour deformation.
One of the arguments made for deforming contour symmetrically
\begin{align}
  \label{eq:4}
  C(t;\beta)_{\text{regulated}}= \Tr\left(\rhoh [W(t),V]^\dagger\rhoh [W(t),V]\right)\,,
\end{align}
or
\begin{align}
  \label{eq:7}
  F(t,t)_{(\frac{1}{2},\frac{1}{4})} = \Tr\left(\rhoq W(t)\rhoq V
  \rhoq W(t)\rhoq V \right)\,,
\end{align}
is that the smearing of the density matrix regulates a short distance
singularity by separating the local operators in imaginary time. If
this were indeed what the smearing should accomplish, then (1) at any
finite value of regulator $\eta$ we should expect the {\it low}-temperature
limit to be universal, and (2)
at any finite temperature $\beta$ in units of the mass $m$ the answer
for the OTOC should diverge as one removes the regulator $|\eta|
\rightarrow \frac{\beta}{2}$. 
The result, however, shows the
opposite. The high-temperature limit is universal, indicating that
this is the regime that is insensitive to the regulator, and, though we do not compute the full OTOC,
the Lyapunov spectrum at fixed $\beta m$ stays finite for any value of
regulator.
This argues strongly that the contour-deformation regulates the IR
rather than the UV. This in fact agrees with Schwinger-Keldysh
theory. There, the ``contour-deformation'' is the introduction of
temperature itself, and this is a well-known IR regulator. 

For results in the literature in perturbative QFTs, this diametrically
opposite interpretation of the contour deformation has little
effect. As in e.g \cite{Stanford2016,Grozdanov:2018atb, Werman:2017abn, Klug:2018ovi}  usually the
focus is on the 
universal high temperature regime. However, for the SYK model, the
focus has often been on the emergent regime at low temperatures. There,
this realization that the contour deformation regulates the IR may imply
that the results are in fact regulator dependent and do not reflect physical
information about the true dynamics. As we will show in
  Section~\ref{sec:SYK}, SYK is special in that its gapless nature and
  the absence of a thermal mass imply contour independence of the
  Lyapunov spectrum even at
  low temperature, extending from the $\beta m \to 0$ regime of weakly coupled field theories.
Before we turn to this, we first address how to obtain the physical
information about the true chaos/scrambling dynamics at low temperatures.

\subsection{Kinetic theory interpretation of the $\alpha$-deformed OTOC}
\label{sec:kinet-theory-interpr}

IR regulators often encode real physical circumstances. The correct
question to ask therefore is which contour properly reflects
physical information of microscopic chaos. In this section we will
argue that this can be 
decided by interpreting  the result of the previous section in terms
of the kinetic theory for many body chaos derived in
\cite{Grozdanov:2018atb}. There, the authors showed that the
computation of the $\alpha=1/2, \sigma =1/4$ OTOC is equivalent to a
Boltzmann-like equation that tracks the time evolution of the gross
energy exchange. 

We briefly review this result. The standard Boltzmann equation describes the time evolution of the single-particle distribution function $f(t,\mathbf{r},\mathbf{p})$,\footnote{Not to be confused with the commutator-squared function defined in Eq. \eqref{eq:f_matrix}.} parametrizing the deviation of the single-particle distribution function from its equilibrium value:
\begin{equation}
f(t,\vec{p}) =\frac{\delta  n(t,\vec{p})}{(1+n(\vec{p}))n(\vec{p})}\,,
\end{equation}
and  $n(\vec{p})$ is the Bose-Einstein distribution. For small deviations from the equilibrium value, the Boltzmann equation can be linearized and, focusing on the 
 homogeneous case, it reads
\begin{equation}
\partial_t f(t,\vec{p}) = -\int\limits_{\vec{l}} \mathcal{L}(\mathbf{p},\mathbf{l}) f(t,\vec{l})\,,
\end{equation} 
where $\mathcal{L}(\mathbf{p},\mathbf{l})$ represents the collision integral. $\mathcal{L}(\mathbf{p},\mathbf{l})$ contains two contributions, namely the \textit{gain} term $R^{\wedge}(\mathbf{p},\mathbf{l})$, counting increase of the density of the phase-space cell, and the \textit{loss} term $R^{\vee}(\mathbf{p},\mathbf{l})$, which accounts for scattering out of the phase-space cell. In terms of these two contributions,  the Boltzmann equation is
\begin{equation}\label{eq:Net}
\partial_t f(t,\vec{p}) = \int\limits_{\vec{l}} \big[
R^{\wedge}(\mathbf{p},\vec{l})-R^{\vee}(\mathbf{p},\vec{l})
\big] f(t,\vec{l}).
\end{equation}
As shown in  \cite{Grozdanov:2018atb}, the Bethe-Salpeter equation of
the symmetrised commutator-squared $C(t;\beta)_{(\half,0)}$ is
equivalent  to considering a Boltzman-like equation where the sign of
the contribution of the true loss term is changed, so that we account for a  gross exchange  rather than a net exchange. More precisely, the gross exchange is given by
\begin{equation}\label{eq:Gross}
\partial_t f^{\mathrm{EX}}(t,\vec{p}) =  \int\limits_{\vec{l}}
\frac{\mathcal{E}[E_{\vec{p}}]}{\mathcal{E}[E_{\mathbf{l}}]} 
\big[R^{\wedge}(\mathbf{p},\vec{l})+R^{\vee}(\mathbf{p},\vec{l})- 4\Gamma_{\vec{l}}\delta(\mathbf{p}-\vec{l})
\big] f^{\mathrm{EX}}(t,\vec{l})\,,
\end{equation}
where $\mathcal{E}[E_{\vec{p}}]=1/\sinh(E_{\vec{p}}\beta/2)$ is an energy-related observable which does not alter the spectrum of the collision integral, as it enters in the form of a similarity transformation. The extra factor $\Gamma_{\vec{l}}$, the self-energy due to the thermal environment, is present to avoid over-counting. It can be understood as follows: 
 $R^{\vee_T}(\mathbf{p},\mathbf{l})\equiv
 R^{\vee}(\mathbf{p},\vec{l})-
 2\Gamma_{\mathbf{l}}\delta(\mathbf{p}-\mathbf{l})$ counts the changes
 in the particle number  $f(t,\mathbf{p})$ due only to processes with
 $\mathbf{p}\neq\vec{l}$. Therefore, changing the sign of $R^{\vee}$
 in Eq. \eqref{eq:Net} would over-count the contribution from the
 bath. If one changes only the sign of the true loss term
 $R^{\vee_T}(\mathbf{p},\mathbf{l})$, the gross exchange is exactly
 given by
 $R^{\wedge}(\mathbf{p},\vec{l})+R^{\vee}(\mathbf{p},\vec{l})-
 4\Gamma_{\vec{l}}\delta(\mathbf{p}-\vec{l})$ \cite{Grozdanov:2018atb}.
The eigenvalues  of the integral operator $\eqref{eq:Gross}$  are
equivalent to those measuring the exponential growth rate of the OTOC,
and thus give the Lyapunov spectrum of the theory.

As the $\alpha$-deformation only changes the rung function in the
Bethe-Salpeter equation, resulting in result \eqref{eq:BS_eta}, it is
immediately recognized that the kinetic equation encoding the late
time behavior of these families of OTOC is modified as follows
\begin{align}
\partial_t f^{\mathrm{EX}}(t,\vec{p})=  \int_\mathbf{l} 
\frac{\mathcal{E}[E_{\mathbf{p}}]}{\mathcal{E}[E_{\mathbf{l}}]}&\Big\{
 \cosh\left[ \eta(E_\p-E_\mathbf{l})\right]
                                        R^{\wedge}(\mathbf{p},\mathbf{l})
                                        \nonumber\\
 &+ \cosh\left[ \eta(E_\p+E_\mathbf{l})\right]
   R^{\vee}(\mathbf{p},\mathbf{l}) \nonumber\\
&- 2\big[\cosh\left[ 2\eta(E_\p)\right]+1\big]\Gamma_{\mathbf{l}}\delta(\mathbf{p}-\mathbf{l})\Big\} f^{\mathrm{EX}}(t,\vec{l})\,.\label{eq:GrossAlpha}
\end{align}

The kinetic equation equivalent of the contour-dependent commutator
square gives us a direct physical
interpretation of what is computed, as we understand each term as
loss, gain and self-energy terms in the microscopic dynamics. The
explicit $\eta=\beta(\alpha-1/2)$  dependence in Eq.  \eqref{eq:GrossAlpha} shows that the
different contours in the $\alpha$-family have a different physical
origin. While for $\eta = 0$ (symmetric regularization) both the
{gain} and  {loss} processes are weighted equally, for other contours
$\eta\neq0$, their relative weight is different.  For none of these values does the
kinetic equation have an obvious natural physical interpretation in terms of gross, net
or otherwise simple exchange dynamics.

On the other hand, the gross exchange equation has been put forward
independently already a long time ago as a measure of microscopic
classical chaos \cite{VanZon2000}. 
This conclusion from the weakly coupled field theory computation above
therefore strongly suggests that, in order to probe dynamical
many-body chaos in QFT, the correct choice for the out-of-time
correlation function is the symmetrically regularized choice with
$\eta=0$. Fortuitously, this is the one that has predominated all the
calculations in the literature, including the derivation of the MSS
bound on chaos \cite{Maldacena2016c}. 
It also means that the naive {\it thermal} expectation value of
the commutator-squared $\Tr[\rho [W(t),V]^2]$ does {\it not} measure
microscopic quantum chaos. One is therefore left with the reversed
question: how does one justify from first principles the symmetrically
regularized commutator-squared as a measure of quantum chaos. We will
return to this question in the last section. First, we will consider the
same question of contour-dependence of the commutator-squared and  its
Lyapunov spectrum for the case of the SYK model.

\section{Contour dependence of the Lyapunov exponent in the SYK
  model}\label{sec:SYK}

One of the research directions where the commutator-squared has had
important impact is in the emergent strongly coupled low energy regime
of the Sachdev-Ye-Kitaev model. The exponential growth of the
symmetrically regularized commutator-squared saturates the MSS bound on
chaos $\lambda_L\leq 2\pi T$; this has given great impetus to the
notion that the SYK model provides a microscopic theory for AdS black
holes.

Now that we know that the commutator-squared and its Lyapunov spectrum
depend on the way the contour is regulated, the natural question on how this affects the insights in the SYK model
arises. We shall first
show that, in contrast to the previous weakly coupled massive QFT
  results, in the SYK model the Lyapunov spectrum is
contour regularization independent.

The SYK Hamiltonian with $q/2$-body interactions is
\begin{align}\label{hami}
\hspace{-1mm}H = i^{q\over 2} \hspace{-4mm}\sum_{1\leq i_1<i_2<\dots<i_q\leq N}\hspace{-4mm} J_{i_1, i_2,\dots, i_q}\ \chi_{i_1}\chi_{i_2} \dots \chi_{i_{q}} \,,
\end{align}
where $\chi_i$ are Majorana fermions so
$
\{ \chi_i, \chi_j \} = \delta_{ij}$ and
the coupling $J_{i_1, i_2,\dots, i_q}$ is a Gaussian-distributed random variable
with zero average and diagonal (i.e. for each $J_{i_1, i_2,\dots, i_q}$
independently) variance
 ${2^{q-1}\over q }{J^2(q-1)!\over N^{q-1}}$
\cite{Maldacena2016}. The fermionic  two-point function
$G(\tau)=-\langle \mathcal{T} \chi(\tau)\chi(0)\rangle$ satisfies the
following averaged Dyson equation in the large-$N$ limit \cite{Maldacena2016}:
\begin{equation}\label{eq:DysonSYK}
G_n^{-1}={-i\omega_n -\Sigma_n }\,,\  \Sigma(\tau)=-J^2 G(\tau)^{q-2}G(-\tau)\,,
\end{equation}
with  $\omega_n=(2\pi/\beta)(n+1/2)$,  $G_n\equiv G(i\omega_n)$ and $\Sigma_n\equiv \Sigma(i\omega_n)$.
In the same way as for weakly coupled QFT, the symmetrical contour regularized commutator-squared $C(t;\beta)_{(\half,0)}$
satisfies a Bethe-Salpeter equation. In the large-$N$ limit, for
arbitrary coupling, the homogeneous part of the BS-equation
governs the large time limit and 
is \cite{Maldacena2016}:
\begin{align}\label{eq:BS}
F(t_1,t_2)=J^2(q-1)\int \dd t_3\dd t_4 G^R(t_{13})G^R(t_{24})\left[G^W(t_{34})\right]^{q-2}F(t_3,t_4)\,,
\end{align}
where $G^R$ and $G^W$ are the retarded and Wightman two-point
functions.

There is now a difference with the
the perturbative QFT approach. As recalled in the previous section,
there the late time approximation also involves a pinching pole
``on-shell'' reduction of the retarded Green's functions. The large $N$
late time limit in SYK, on the other hand, is a conformal field theory with no on-shell
particle-like excitations. There is no natural simplification of the
retarded SYK Green's functions in this limit. In contrast to the
perturbative QFT
solution, the full large $N$ Green's functions are obtained by analytically continuing the Dyson equation Eq. \eqref{eq:DysonSYK} to real time and solving these equations numerically with an iterative procedure \cite{Garcia-Garcia2018b}.\footnote{
As we are using the symmetric regularization,
$G^W(\omega)=\rho(\omega){e^{-\omega{\beta\over2}}\over 1+e^{-\beta
    \omega}}$ is the Wightman function with operators separated by
$i\beta/2$.} Then one solves the SYK BSE
Eq.\eqref{eq:BS} by making the explicit ansatz $F(t_1,t_2) = e^{\lambda_L(t_1+t_2)/2}f(t_{12})$ and rewriting it as an integral eigenvalue equation in frequency space:
\begin{align}
&f(\omega'){=}(q-1)J^2\left|G^R\left(\omega'{+}i{\lambda_L\over 2}\right)\right|^2 \hspace{-2mm}\int\hspace{-1mm} {\dd\omega\over 2\pi}
g_{lr}(\omega'{-}\omega)f(\omega)\,,\label{eq:BS_freq}\\
&g_{lr}(\omega)\equiv\int \dd  te^{i\omega t}G^W(t)^{q-2}\,.\notag
\end{align}
One finally (numerically or analytically) searches for which value of
$\lambda_L$ the kernel has an eigenvector with eigenvalue 1 \cite{Maldacena2016}.

We can now ask how the subtly different SYK computation of its
  Lyapunov spectrum depends on the contour.
As in the perturbative QFT of the Sec. \ref{sec:weak}, the only place the
contour regularization shows up is in the Wightman functions.\footnote{This can also be seen explicitly by analytically continuing Eq. \eqref{eq:BS} $t_1\to t_1+i\beta\left(\alpha-\quarter\right)\,,\ t_2\to t_2+i{\beta\over 4}\,, t_3\to t_3+i\beta\left(\alpha-\half\right)\,,\ t_4\to t_4$.}
 Instead
of parametrizing with respect to the $\alpha=0$ Wightman function, let us
parametrize with respect to the $\alpha=1/2$ Wightman function:
\begin{equation}
G^\eta(\omega)= e^{\eta\omega} G^{\eta=0}(\omega)\,.
\end{equation} 
The Bethe-Salpeter equation \eqref{eq:BS} for the commutator
square in frequency space for
arbitrary $\alpha$-deformed contour is then the same as before, but with a modified kernel $\tilde g_{lr}(\omega)$:
\begin{align} 
&\tilde f(\omega'){=}
(q-1)J^2\left|G^R\left(\omega'{+}i{\lambda_\alpha\over 2}\right)\right|^2 \hspace{-2mm}\int\hspace{-1mm} {\dd\omega\over 2\pi}
\tilde g_{lr}(\omega'{-}\omega)\tilde f(\omega)\,,\label{eq:BS_other_reg_tmp}\\
&\tilde g_{lr}(\omega)\equiv\int \dd  te^{i\omega t}G^\eta(t)^{q-2}\,,\quad 
G^\eta (t) =\int \dd  \omega e^{-i\omega t}G^\eta(\omega)\,.
\end{align}
We evaluate the modification in the kernel $\tilde g_{lr}(\omega)$,
compared to the original kernel $g_{lr}(\omega)$, by using the convolution of the Wightman functions:
\begin{align} 
&\tilde g_{lr}(\omega)=\int \dd  te^{i\omega t}G^\eta(t)^{q-2} = 
\int \dd \omega_1\dots \dd \omega_{q{-}3} 
G^\eta\big(\omega_1\big)
G^\eta\big(\omega_2\big)\dots 
G^\eta\big(\omega-\omega_1-\dots -\omega_{q{-}3}\big) \,,\nonumber
\end{align}
and substituting $G^\eta(\omega)=e^{\omega \eta}G^W(\omega)$ in each term inside the integral:
\begin{align} 
\tilde g_{lr}(\omega)&{=}\hspace{-4mm}
\int\limits_{ \omega_1,\dots, \omega_{q{-}3} }\hspace{-4mm}
e^{\eta \omega_1}G^\eta\big(\omega_1\big)
~ e^{\eta \omega_2}G^\eta\big(\omega_2\big)\dots 
e^{\eta (\omega-\omega_1-\dots -\omega_{q-3})} G^\eta\big(\omega-\omega_1-\dots -\omega_{q{-}3}\big) \nonumber\\
&{=}~ e^{\eta\omega}\int \dd te^{i\omega t} G^W(t)^{q-2}=
e^{\eta\omega}g_{lr}(\omega)\,.\nonumber
\end{align}
Therefore, Eq. \eqref{eq:BS_other_reg_tmp} reduces to 
\begin{align} 
&\tilde f(\omega'){=}
(q-1)J^2\left|G^R\left(\omega'{+}i{\lambda_\alpha\over 2}\right)\right|^2 \hspace{-2mm}\int\hspace{-1mm} {\dd\omega\over 2\pi}
e^{\eta(\omega'{-}\omega)}
g_{lr}(\omega'{-}\omega)\tilde f(\omega)\,,\label{eq:BS_other_reg}
\end{align}

The crucial
difference with weakly coupled QFT is that, because of the gapless
nature of SYK even at finite temperature and the absence of a pinched
pole on-shell condition, the product of retarded Green's functions
remains a smooth function and not a distribution. This allows one to
reabsorb the contour dependence with the redefinition $\tilde
f(\omega')\to \tilde e^{-\eta \omega'}f(\omega')$.\footnote{We
    thank Subir Sachdev and Yingfei Gu for emphasizing that this
    should be the case.} In the late time 
SYK-BSE, this acts as a similarity transformation on the kernel, and its
eigenvalues equal to its Lyapunov spectrum is therefore contour
independent. Note again that the OTOC is still contour dependent, yet the
Lyapunov exponent is independent of the regularization. As mentioned
before, this can be qualitatively understood from the weakly coupled
case, where in the massless case the dependence of the contour on the
Lyapunov exponent vanishes.

To check the solution obtained with the BSE,  we now consider the two regimes of the SYK model 
where some analytical control is possible:
the strong coupling limit $\beta J\gg 1$ and the large-$q$ limit.

\subsection{Study of the OTOC in SYK in the strongly coupled limit: conformal limit analysis }

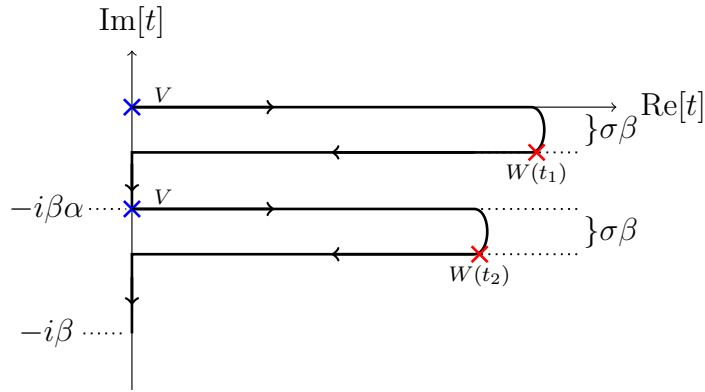
\begin{figure}[t]
 \centering
\begin{tikzpicture}[scale=0.75][h]
\hspace{8mm}
\begin{scope}
\draw[line width=0.15mm,->] (-1,-3) to (-1,3) ;
\draw[line width=0.15mm,->] (-1,2) to (7.5,2) ;
\node at (-1,3.5) {$\mathrm{Im}[t]$}; 
\node at (8.5,2) {$\mathrm{Re}[t]$};
\draw[line width=0.35mm] (-1,2) to (6,2) to [out=0, in =0] (6,1.2) to (-1,1.2) to (-1,0.2) to (5,0.2) to [out=0, in =0] (5,-0.6) to (-1,-0.6) to (-1,-2);
\draw (-1,2) node[cross=4pt,line width=0.35mm,blue] {};
\draw (-0.45,1.9) node[above] {$\scriptstyle V$};


\draw (6.1,1.2) node[cross=4pt,line width=0.35mm,red] {};
\draw (6.1,1.2) node[below] {$\scriptstyle W(t_1)$};

\node at (-2.5,0.2) {$-i {\beta\alpha}$}; 
\draw[dotted,line width=0.25mm] (-1.75,0.2) to (-1.1,0.2) ;
\draw (5.1,-0.6) node[cross=4pt,line width=0.35mm,red] {};
\draw (5.1,-0.6) node[below] {$ \scriptstyle W(t_2)$};


\draw (-1,0.2) node[cross=4pt,line width=0.35mm,blue] {};
\draw (-.45,0.1) node[above] {$ \scriptstyle V$};

\node at (-2.5,-2) {$-i\beta$}; 
\draw[dotted,line width=0.25mm] (-1.85,-2) to (-1.1,-2) ;
\draw[line width=0.35mm, ->] (-1,2) to (1.5,2) ;
\draw[line width=0.35mm, ->] (5,1.2) to (2.5,1.2) ;
\draw[line width=0.35mm, ->] (-1,0.2) to (1.5,0.2) ;
\draw[line width=0.35mm, ->] (5,-0.6) to (2.5,-0.6) ;
\draw[line width=0.35mm, ->] (-1,1) to (-1,0.5) ;
\draw[line width=0.35mm, ->] (-1,-1) to (-1,-1.5) ;

\draw[dotted,line width=0.25mm] (5,0.2) to (6.9,0.2) ;
\draw[dotted,line width=0.25mm] (5,-0.6) to (6.9,-0.6) ;
\node at (7.4,-0.2)  {$\} { \sigma\beta}$}; 

\draw[dotted,line width=0.25mm] (5,1.2) to (6.9,1.2) ;
\node at (7.4,1.6) {$\} { \sigma\beta}$}; 

\end{scope}

\end{tikzpicture}
\vspace{-2mm}
	\caption{Extended Schwinger-Keldysh contour  corresponding to the two-parameter OTOC $\Tr[\rhoS{\alpha} W(t_1+i\sigma\beta)V\rhoS{1-\alpha}W(t_2+i\sigma\beta) V ]$. }\label{fig:SK_alpha_delta}

\end{figure}

In the strongly-coupled regime $\beta J\gg1$ of the SYK model, where conformal symmetry emerges asymptotically, the OTOC may also be computed analytically by studying the spectrum of the Casimir operator. More specifcally, for $\beta J\gg1$ the eigenvectors of the Casimir operator, with eigenvalue $h(h-1)$, are also eigenvectors of the Euclidean kernel of the Bethe-Salpeter equation  \cite{Maldacena2016}. In this regime, the kernel of the Bethe-Salpeter equation is:
\begin{align}
K_c(\tau_1,\tau_2;\tau_3,\tau_4) &\propto
\frac{\sign(\tau_{13})\sign(\tau_{24})}{|\tau_{13}|^{2\Delta}|\tau_{24}|^{2\Delta}|\tau_{34}|^{2-4\Delta}}\,,\quad \Delta=1/q\,,\notag
\end{align}
where  the eigenvalues of $K_c$  depend on $q$ and $h$.  Moreover, the allowed values of $h$  are constrained, because the Bethe-Salpeter equation for the OTOC selects  the eigenvalue unity of $K_c$. For $q=4$, the leading contribution to the OTOC turns out to be $h=2$ and is given by \cite{Maldacena2016}:
\begin{align}\label{eq:conformalK}
\frac{\CF(\theta_1...\theta_4)}{G(\theta_{12})G(\theta_{34})} &= \frac{6\alpha_0}{\pi^2\alpha_K}\beta \mathcal{J}\sum_{|n|\ge 2}\frac{e^{in(y'-y)}}{n^2(n^2-1)}\bigg[\frac{\sin \frac{nx}{2}}{\tan \frac{x}{2}} - n \cos\frac{nx}{2}\bigg]\bigg[\frac{\sin \frac{nx'}{2}}{\tan \frac{x'}{2}} - n \cos\frac{nx'}{2}\bigg]\notag\\
x=\theta_{12}&\hspace{20pt}x' = \theta_{34}\hspace{20pt}y=\frac{\theta_1+\theta_2}{2}\hspace{20pt}y' = \frac{\theta_3+\theta_4}{2}\,,
\end{align}
where $\theta$ is the rescaled Euclidean time $\theta =\tau/\beta$.
This equation must now be analytically continued to real time by choosing the operator insertions. We consider the contour shown in Fig. \ref{fig:SK_alpha_delta}, which allows us to consider both the $\sigma$- and $\alpha$-families simultaneously.

More specifically, we choose
\begin{align}
\theta_1=i{2\pi t\over\beta}+2\pi(\sigma+\alpha)\,,
&\hspace{20pt}
\theta_2=i{2\pi t\over \beta}+2\pi\sigma\,,\hspace{20pt}
\theta_3=2\pi\alpha\,,\hspace{20pt}
\theta_4=0\,.\notag
\end{align}
In terms of $x,x',y,y'$, we have:
\begin{align}
x=x'=2\pi\alpha\,,&\hspace{20pt}y=i{2\pi t\over \beta}+2\pi\sigma+\pi\alpha\,,\hspace{20pt}
y' = \pi\alpha.\label{eq:choice_times}
\end{align}
In order compute Eq. \eqref{eq:conformalK} explicitly,  we set $x=x'$, sum over $n$ and then substitute Eq. \eqref{eq:choice_times}	to get:
\begin{align}
\frac{\CF(t)}{G(2\pi\alpha)G(2\pi\alpha)} &\propto
\frac{1}{2}-
\frac{\pi}{4}  \left\{2 \pi  \cot ^2(\pi  \alpha ) 
\left[(\alpha -1) \alpha +\sigma +i {t\over\beta}\right]+
  (4 \alpha -2) \cot (\pi  \alpha )\right.\notag\\
&\left.+i \csc ^2(\pi  \alpha ) \sinh 
\left(\frac{2\pi t}{\beta }-2 i\pi \sigma\right) 
\right\}\notag\,,
\end{align}
which for large $t$ behaves as: 
\begin{align}\label{eq:Ft_conf}
\frac{\CF(t)}{G(2\pi\alpha)G(2\pi\alpha)} &\propto
-i{\pi^2\over4} \csc^2(\pi\alpha)e^ 
{-2 i\pi \sigma}
e^ 
{\frac{2\pi t}{\beta }}\,.
\end{align}

We first note that $\CF(t)$ is symmetric over $\alpha\to 1-\alpha$, as
expected. Second, the long-time regime is controlled by a growth rate
given by $2\pi/\beta$, independent of the distance between the forward
branches $\alpha$. This confirms the contour dependence of the
  OTOC while the Lyapunov spectrum is nevertheless independent.
  The same contour dependence appears in 2-dimensional CFTs in the large central charge limit, where the OTOC is dominated by the Virasoro identity block \cite{Roberts2014}. In this case the contour-dependence always enters as prefactors of the $e^{2\pi t\over \beta}$ as in Eq. \eqref{eq:Ft_conf}.

\subsection{Study of the OTOC in SYK in the limit of large interaction
  order}

In the SYK model, analytical control is also possible when one 
increases the order of the interaction in the Hamiltonian  \eqref{hami}, which is
set by $q$ \cite{Maldacena2016}. Here we consider the
calculation of the Lyapunov exponent in SYK in the large-$q$ limit,
and show that it is also contour independent.

We start with the two-point function in Euclidean signature in the
large-$q$ expansion 
\cite{Maldacena2016}:
\begin{align}
\label{eq:largeq:expansion}
G(\tau) \underset{q\gg 1}{=} \frac{1}{2}\sgn(\tau)\left(1+\frac{1}{q} g(\tau)+O(q^{-2})\right)\,,
\end{align}
where $g(\tau)$ is obtained by inserting the above ansatz in the saddle point equation for the two-point function. This gives the equation
\begin{align}
\label{eq:largeq:eom}
\partial_{\theta}^{2}g = 2 (\beta  \JJ)^2e^{g(\theta)}\,,
\end{align}
\noindent where $\theta = \tau/\beta \in [0,1)$ and $\JJ^2=q2^{1-q}J^2$, and with   boundary conditions $g(0)=g(1)=0$. The solution of Eq. \eqref{eq:largeq:eom}	is 
\begin{align}
\label{eq:large_q_sol}
e^{g_{(0)}(\theta)} = \left[\frac{\cos{\frac{\pi\nu}{2}}}{\cos\left[\pi\nu\left(\frac{1}{2}-\theta\right)\right]}\right]^2,	&&\beta  \JJ = \frac{\pi\nu}{\cos{\frac{\pi\nu}{2}}}\,,
\end{align}
with  $\nu\in [0,1]$ parametrising the flow from weak $\beta J\sim0$ coupling ($\nu\sim0$), to strong coupling  $\beta J\gg1$ ($\nu\sim1$).
The analytic continuation to real time reads
\begin{align}
G^R(t)&=\theta(t)\left[G(\tau\to it+\epsilon) -G(\tau\to it-\epsilon)\right]=\theta(t)+\OO(1/q),\notag\\
G^{(\alpha)}(t)&=G(\tau\to it+\alpha\beta)\,,\label{eq:largeq_cont}
\end{align}
for $\alpha=1/2$, and $G^{(\alpha)}(t)$ gives the Wightman function with operators separated by $i\beta/2$.
Instead of working in frequency space with Eq. \eqref{eq:BS_other_reg}, we work in the time domain and use the following simplification for large $q$:
\begin{align}
J^2(q-1)G(\tau)^{q-2} \underset{q\gg1}{\simeq} J^2q 2^{2-q}\sgn(\tau)^{q-2}e^{g(\tau)} =2\JJ^2 e^{g(\tau)}\,.\label{eq:largeq_simp}
\end{align}
Therefore, using Eqs. \eqref{eq:largeq_cont}	and \eqref{eq:largeq_simp},	the kernel of the Bethe-Salpeter equation, Eq. \eqref{eq:BS}, for large-$q$ is
\begin{align}
K^{(\alpha)}(t_1,\dots,t_4)&= J^2(q-1)G^R(t_{13})G^R(t_{24})G^\alpha(t_{34})^{q-2}\\
&\underset{q\gg1}{\simeq} \theta(t_{13}) \theta(t_{24}) 2\mathcal{J}^2 e^{g(\tau\to it_{34} +\beta \alpha)}\nonumber\\
&= \theta(t_{13}) \theta(t_{24}) 
{2\pi^2\nu^2\over \beta^2 \cosh^2\left({\pi\nu\over \beta}(t_{34}+i\eta)\right)}\,,\qquad \eta = \beta(\alpha-1/2)\,,\nonumber
\end{align}
where $0<\nu<1$ and we take $0<\alpha<1/2$.
Finally, we use $\partial_t \theta(t)=\delta(t)$ to simplify the Bethe-Salpeter equation
\begin{align}
&\partial_{t_1}\partial_{t_2}\left[F(t_1,t_2)=\int \dd t_3\dd t_4 K^{(\alpha)}(t_1,\dots,t_4)F(t_3,t_4)\right],\nonumber\\
&\partial_{t_1}\partial_{t_2}F(t_1,t_2)= 2\mathcal{J}^2 e^{g(\tau\to it_{12} +\beta \alpha)}F(t_1,t_2)\,.
\end{align}
Making the ansatz $F(t,t')=e^{\lambda_L {t+t'\over2}}f(t-t')$:
\begin{align}
\left[\partial_{t_{12}}^2+2\mathcal{J}^2e^{g(\tau\to it_{12} +\beta \alpha)}\right]f(t_{12}) =&{\lambda_L^2\over4}f(t_{12}),\nonumber\\
\left[\partial_{t_{12}}^2+{\pi^2\nu^2\over\beta^2} {2\over \cosh^2\left({\pi\nu\over \beta}(t_{12}+i\eta)\right)}\right]f(t_{12}) =&{\lambda_L^2\over4}f(t_{12}),\nonumber\\
\left[-\partial_{y}^2- {2\over \cosh^2\left(y+i\frac{\pi\nu}{\beta}\eta\right)}\right]f(y) =&-\left({\lambda_L \beta\over2\pi\nu}\right)^2f(y)\,.\nonumber
\end{align}
This is the Schr\"{o}dinger equation with a \textit{complex} P\"oschl-Teller potential, which has a boundstate, $f(y)=\sqrt{\tan(\pi\nu\eta/\beta)\over 4\eta\cos(\pi\nu\eta/\beta)} \sech(y+i\frac{\pi\nu}{\beta}\eta)$, with real eigenvalue $E=-1$. 
The value of the eigenenergy gives the value of the Lyapunov exponent
$\lambda_L=2\pi\nu/\beta$, which is independent on the contour
parameter $\alpha$.

The large $q$ analysis allows us a qualitative insight into the
  role of the gapless nature of SYK by taking a closer look to the
  SYK-BSE Eq. \eqref{eq:BS_other_reg}. Take the kernel of the
Bethe-Salpeter equation in the regime where conformal symmetry is only
weakly broken $\beta J\gg1$. In this regime, the symmetric
  $\eta=0$  Wightman function is $G^W(t) =b \left[\frac{\pi
  }{\beta\cosh\frac{\pi t}{\beta}}\right]^{2/q}$, $b^q =  \left({
    \half -{1\over q} } \right){    \tan( \pi/q)}/(J^2\pi)$
\cite{Maldacena2016}.  Consequently, the $\eta=0$ kernel $g_{lr}(\omega)$  is
\begin{align}
g_{lr}(\omega'-\omega)=\int \dd t e^{i\omega t} G^W(t)^{q-2}=b^{q-2}\left(\pi\over \beta\right)^{2-4/q}{2^{1-4/q}\over \Gamma\left(2-{4\over q}\right)}\left|\Gamma\left(1-{2\over q}-i{\beta (\omega'-\omega)\over2\pi} \right)\right|^2.\notag
\end{align}
Using the identity $|\Gamma(a+ib)|^2 =
\Gamma(a)^2\prod_{k=0}^{\infty}\frac{1}{1+b^2/(a+k)^2}$, one
immediately sees that this kernel is strongly peaked around the origin
$\beta(\omega'-\omega) \rightarrow 0$. On the other hand, changing the regularization
changes the kernel by an overall factor
$e^{(\alpha-1/2)\beta(\omega'-\omega)}$. Thus, as the integral in
Eq. \eqref{eq:BS_other_reg} is dominated by $\omega'\sim \omega$, the
dependence on the contour proportional to
$e^{\eta(\omega'{-}\omega)}\sim1$ essentially drops out.

In a theory with gapped excitation, on the other hand, one can
  see for the case of the matrix model by numerically inspecting
  the expression of the on-shell kernel \eqref{eq:Lyapunov_weak} and
  \eqref{eq:IKernel}, that 
  the kernels are peaked around the value of the gap. This gives a
  contribution of the order $e^{\eta \Delta_{gap}}$. Physically it may
  be seen as a consequence of a combination of an on-shell
  particle and anti-particle process that dominates the kernel.

\section{The Lyapunov spectrum and the Loschmidt
  echo}\label{sec:Loschmidt}

In  previous sections we have seen how the regularization dependence can afflict the commutator-squared. This
shows that without more detailed specification one cannot directly relate this quantity to an observable
that can be measured in experiments. We have also shown that the
regularization dependence is dominant in the IR rather than the
UV. This is analogous to Schwinger-Keldysh theory where contour
dependence is related to the temperature, and the latter is a well
known IR regulator. IR regularization issues are usually not solved by
counterterms and renormalization. Instead they often encode physics on
their own. This suggests that a way to resolve the regulator
dependence is to {\it define} which member of family of
``regularized'' correlation functions computes a proper physical
observable. The weakly coupled QFT result, through the mapping of
the commutator-squared to a kinetic equation, 
indicates that the symmetrically regularized commutator-squared is the
correct one. 

Fortuitously this is the one almost exclusively studied in
the literature and the one for which the MSS bound on chaos is
derived. Nevertheless, one would like to understand from first
principles why the symmetrized contour is an appropriate physical
observable. The first attempt construction in the introduction points
to the thermally averaged commutator-squared instead. 
In this section we show that the symmetrized commutator-squared follows
directly from an alternative measure of chaos, which is related to standard measurements of information spreading: the Loschmidt echo. This quantity contains not only the commutator-squared but also higher-order out-of time correlation functions.
The Loschmidt echo and related quantities have been used in the context of quantum chaos for a long time \cite{Pastawski1995,Jalabert2001,Prosen2002,Prosen2002a,Haug2005}.
Therefore, it is not surprising that the OTOCs may be extracted from echo spectroscopy as proposed in \cite{Zhu2016,Yao2016,Swingle2016,Kurchan2018} and measured experimentally in \cite{Garttner2017}. More recently, the connection between the OTOC and the Loschmidt echo at infinite temperature was also explored in \cite{Bin2019}.

\subsection{Loschmidt echo}

The Loschmidt echo is based on a old thought experiment trying to
disprove the irreversibility inherent in Boltzmann's equations by
imaging a dynamical system where at time $t$ after $t_0=0$ one reverses all
velocities and compare the resulting state at time $2t$ with the
original state. Microscopically the answer is  of course identical,
but supposing one makes a tiny ``erroneous'' perturbation at the time when one
reverses all velocities, one immediately sees that in a chaotic
non-integrable system the resulting state will be exponentially
different from the original state. 

This thought experiment can be directly mapped to a quantum quench
experiment. One evolves a quantum state forward in time for a time
$t$, perturbs it with an instantaneous quench $e^{i\delta W} $,
evolves backward for the same time $t$ and projects onto the original
state,
\begin{equation}\label{eq:Loschmidt}
M(t)\equiv \langle \psi| e^{iHt}e^{i\delta W} e^{-iHt}|\psi \rangle\,.
\end{equation}
For a generic initial state, the echo will have a universal late time
exponential fall off independent of the type of quench $W$ that encodes the lack of overlap between the initial
and final state,
\begin{align}
  \label{eq:11}
  M(t) = \langle \psi_{\text{initial}}|
  \psi_{\text{Loschmidt}}(2t)\rangle \sim e^{-\lambda t}.
\end{align}
The Lyapunov exponent $\lambda$ is then a property of the system
characterized by its Hamiltonian $H$ alone.

The Loschmidt echo is the expectation value of a complex
operator. To avoid phases one often takes the absolute value squared,
which is known as the fidelity \cite{Prosen2003}
\begin{align}
  \label{eq:10}
  F(t) \equiv \left|\langle \psi| e^{iHt}e^{i\delta W} e^{-iHt}|\psi
  \rangle\right|^2 \sim \left|\frac{1}{1+\delta e^{\lambda
  t}}\right|^2 \stackrel{t\rightarrow \infty}{\sim} e^{-2\lambda t}\,.
\end{align}
The intermediate step is a well-known result from Jalabert and Pastawski \cite{Jalabert2001}.
A second practical step with an eye on experiment is to consider the
fidelity for an ensemble of states, rather than a single state. Choosing the thermal ensemble one has
\begin{align}
  \label{eq:12}
   F(t) &\equiv \left|\frac{1}{Z_{\beta}}\sum_{\psi} e^{-\beta E[\psi]}\langle \psi| e^{iHt}e^{-i\delta W} e^{-iHt}|\psi
  \rangle\right|^2 
          \nonumber \\
  &= \left|\text{Tr}\, \rho e^{iHt}e^{-i\delta W} e^{-iHt}\right|^2
    \nonumber \\
  & = \text{Tr}\, \rho e^{iHt}e^{-i\delta W} e^{-iHt} \rho
    e^{iHt}e^{i\delta W} e^{-iHt}  + {\cal O}\left({1/t}\right) \nonumber \\ 
 &= \text{Tr}\,\rho e^{-i\delta W(t)}\rho e^{i\delta W(t)}.
\end{align}
Defining
$e^{-iHt}\rho e^{iHt} \equiv X$ and $e^{-i\delta W} e^{-iHt}\rho
e^{iHt} e^{i\delta W} = Y$, the fidelity $F(t)$ above is a specific case of
the more general operator fidelity ${\cal F} = \Tr X^{\dagger}
Y$ applied to density matrices as operators.\footnote{The operator fidelity is a weaker version of state
  fidelity encoding the notion of how close a state is to a maximally
  entangled one \cite{Bennett1993} or, if referring to teleportation,
  it quantifies the quality of the teleportation that can be achieved
  with the given state \cite{Jepsen1996}.  The state fidelity between
  two quantum states given by the density matrices $\rho_0$ and
  $\rho_1$  equals \cite{Uhlmann1976,Jozsa1994}:
\begin{equation}\label{eq:state_fidelity}
F(\rho_0,\rho_1) \equiv \Tr\sqrt{\rho_1^{1/2}\rho_0\rho_1^{1/2}}\,.
\end{equation}} 
Three remarks are in order. (1) In the intermediate step we used that
the {\it leading} Lyapunov decay rate in $t$ is the same when computed via
$\left|\text{Tr} \rho e^{-i\delta W(t)}\right|^2$ or $\text{Tr} \rho
  e^{-i\delta W(t)}\rho e^{i\delta W(t)}$.
(2) Naively, as the late time
Lyapunov exponent of interest is a property of the system and not of
the initial state, the
averaging should not matter. However, it is well known from classical
dynamical systems that the late time behavior of an ensemble of
classical trajectories is governed by Policott-Ruelle decay, rather
than the microscopic exponential growth. Even though these are
qualitatively related in weakly coupled theories, they are not quantitatively the same
\cite{Grozdanov:2018atb}.
(3)
Note both the symmetrized appearance of the density
matrix, and the fact that the cumulative power of the density matrix
is 2. Computed through a path-integral this implies that the
periodicity in Euclidean time is twice the inverse temperature $\beta
= 2/T_{\text{phys}}$.

To connect with the commutator-squared, we expand to second order in $\delta$
\begin{align}
  \label{eq:13}
  F(t) &=\Tr\rho^2+ \text{Tr}\,\rho(-\delta^2W(t)^2) \rho + \text{Tr} \,
         \rho (\delta W(t)) \rho \delta W(t) \nonumber \\
& =\Tr \rho^2 +\frac{\delta^2}{2}\text{Tr}\,[\rho, W(t)][\rho,W(t)]  + \ldots 
\end{align}
with the difference that the density matrix itself takes the role of
the operator $V(0)$. The second time dependent term, the density-matrix commutator-squared, is a
variant of the Wigner-Yanase-Dyson skew information.
\begin{equation}
I_\alpha(\rho,A) \equiv \half \Tr[i[\rho^{2\alpha},A^\dagger](i[\rhoS{2-2\alpha},A])]\,, 0\leq\alpha\leq1\,,
\end{equation}
for the symmetric value $\alpha=1/2$ \cite{Wigner1963}. Writing out
the symmetric case for hermitian A, 
\begin{align}
  \label{eq:14}
  I_{\frac{1}{2}}(\rho,A) = \left(\Tr \rho A\rho A - \Tr \rho A A\rho\right)
\end{align}
and replacing the thermal density matrix $\rho$ with a pure state
density matrix,
\begin{align}
  \label{eq:15}
  I_{\frac{1}{2}}(|\psi\rangle\langle\psi|,A) = -(\langle A^2\rangle -
  \langle A\rangle^2),
\end{align}
one can recognize that the WYD skew information is an extension of the variance for pure
states to mixed states. If, by the same argument as above, one may
assume that it is dominated by some largest eigenvalue $ \Tr \rho
A\rho A  \sim (\Tr \rho A)^2$, it computes something akin to the
(largest eigenvalue) variance for
the operator ${\cal O} = \rho A$. In that sense it is again natural that the
density matrix appears with cumulative power 2. Put differently, in
computing the WYD skew information the periodicity in Euclidean time
is twice the inverse temperature $\beta = 2/T_{\text{phys}}$. 

However, this is not yet the commutator-squared we are interested
in. A guess might be the case where the thermal density matrix is rotated
by a small similarity transformation $\rho = e^{iV}\rho_0e^{-iV}$. This is equivalent to an instantaneous quench by $V$ at time $t=1$. 
Then in the limit of small $\delta$ the late time  fidelity equals
\begin{align}
  \label{eq:16}
  F(t) &= \text{Tr}\, \rho e^{iHt}e^{-i\delta W} e^{-iHt} \rho
    e^{iHt}e^{i\delta W} e^{-iHt}  \\
& = \text{Tr}\, \rho_0 e^{iV(0)}e^{-i\delta W(t)} e^{-iV(0)} \rho_0
    e^{iV(0)}e^{i\delta W(t)} e^{-iV(0)} \nonumber \\
& = \Tr \rho_0^2+\frac{\delta^2}{2}\Tr[\rho_0, e^{iV(0)}W(t)
  e^{-iV(0)}][\rho_0,e^{iV(0)}W(t) e^{-iV(0)}] + \ldots \nonumber \\
& = \Tr \rho_0^2 + \delta^2 \left(\Tr \rho_0 e^{iV(0)}W(t)
  e^{-iV(0)}\rho_0e^{iV(0)}W(t)
  e^{-iV(0)} - \Tr \rho_0 e^{iV(0)}W(t)^2
  e^{-iV(0)}\rho_0\right). \nonumber
\end{align}
The first and the last term can never give an OTOC;
ignoring those, one has in the limit of small $V$
\begin{align}
\label{eq:16b}
F(t)  = \text{TOC} +{\delta^2} &\left(\Tr \rho_0W(t)\rho_0 W(t)
  +2\Tr\rho_0W(t)\rho_0[V,W(t)]  + \right.\nonumber\\
& \left.+\Tr \rho_0W(t) \rho_0[V,[V,W(t)]] + \Tr
  \rho_0[V,W(t)]\rho_0[V,W(t)] + \ldots \right).
\end{align}
The two terms of order $\delta^2$ in the first line are also TOC. The terms on the second line contain the
symmetric commutator-squared and a second term which is also an OTOC but
on a different contour.\footnote{ Note that, at higher orders in $\delta$,  the fidelity contains higher-order correlation functions, which  are still represented by a  Schwinger-Keldysh contour with only two folds but with multiple insertions of operators. These correlation functions differ from higher-point OTOCs  in Schwinger-Keldysh contours with more than two folds \cite{Haehl2018}. The latter correspond to multiple repetitions of the Loschmidt experiment and, consequently, the largest growth rate is simply a multiple of the  Lyapunov exponent of the 4-point function OTOC corresponding to a single repetition of the Loschmidt experiment.  }
As
we know by now, generically the Lyapunov behavior of this other OTOC
will be different. This is not yet the answer.

Tracing the origin of Eq. \eqref{eq:16b}, it is easy to see how the
fidelity and the symmetrized commutator-squared are
related. Eq. \eqref{eq:16b} follows from taking the long time limit and
then taking $V$ and $\delta W$ small in the fundamental definition of
the ensemble averaged fidelity --- the first line of
Eq. \eqref{eq:12}. If, however, we take the limit of $V$ and
$\delta W$ small, with $\rho = e^{iV}\rho_0e^{-iV}$, the ensemble averaged fidelity equals
\begin{align}
  \label{eq:17}
  F(t) &= \left|\Tr e^{iV(0)}\rho_0e^{-iV(0)}e^{-i\delta
         W(t)}\right|^2 \nonumber \\
  &= \left|\Tr \rho_0 \left(1 - i\delta W(t) -\frac{\delta^2}{2}W(t)^2
  -\delta [V,W(t)] +\ldots \right)\right|^2 \nonumber \\
 & = \left|\Tr\rho_0\right|^2 -\delta\Tr\rho_0\Tr\rho_0[V,W(t)]
   -\delta\Tr\overline{\rho_0[V,W(t)]}\Tr\rho_0 \nonumber \\
  & ~~~+\delta^2 \Tr\overline{\rho_0[V,W(t)]}\Tr\rho_0[V,W(t)].
\end{align}
We now use the late time
approximation, where we assume that $\rho_0[V,W(t)]$ is dominated by
an eigenvalue Eig$(\rho_0[V,W(t)])\sim
e^{\frac{1}{2}(\lambda+i\phi) t}$. In that limit,
the middle two terms give a strongly oscillatory contribution, which
is hard to measure. We therefore ignore it. As to the last
term in Eq.\eqref{eq:17}, there the late time limit allows us to make
again the approximation
\begin{align}
F(t) &= \ldots + \Tr\overline{\rho_0[V,W(t)]}\Tr\rho_0[V,W(t)] \nonumber\\
&= \ldots +\Tr\overline{\rho_0[V,W(t)]}\rho_0[V,W(t)] + {\cal O}(1/t).
\end{align}
We recognize precisely the
symmetrized commutator-squared with one already noted difference. The
cumulative power of the density matrix is 2. This implies that the
connection between the periodicity
in Euclidean time and the physical temperature differs with a factor
two compared to what the naive smearing procedured assumes: $\beta
=2/T_{\text{phys}}$. In particular this means the proper MSS bound on
chaos should read $\lambda \leq \pi T_{\text{phys}}$. 

The above is a strong argument that the natural observable which measures the
symmetrized commutator-squared is the Loschmidt echo in the limit of
small quenches {first} and late time {subsequent} with the sublety
that $\beta = 2/T_{\text{phys}}$.

\section{Conclusion}

In this article we have explored the role of the regularization scheme of the  commutator-squared and of the OTOC. Quantum chaotic systems may display an exponential growth parametrized by a quantum Lyapunov exponent which is bounded by above $\lambda \leq 2\pi k_B T/\hbar$ \cite{Maldacena2016c}. 
The  proof of this bound involves regularising the OTOC  
by thermally spreading the operators. Purportedly, this is done to
regulate short distance singularities and any physical property of a
system should be independent of the short distance regularization scheme. 

Here, we have shown that for those regularizations consisting on a
contour with a $i\beta/2$ separation between the forward branches,
shifting the backwards branches induces a change in the decoherence
factor, defined as the prefactor of the sum of the OTOCs
\cite{Kitaev2017}. Therefore, the decoherence factor cannot be a
physical quantity as previously suggested. On the other hand,  the
Lyapunov exponent is the same for all of these contours, suggesting
that indeed it may be measurable. 

However, we have then shown that for a different choice of contours,
where the separation between the forward branches is changed, the
Lyapunov spectrum  also depends on the contour  chosen. While the
contour dependence of the commutator-squared has been mostly overlooked in the
literature, it is not surprising that this is the case. Similarly to
the Wightman function, the commutator-squared involves operators
inserted on forward and backward branches of the Schwinger-Keldysh
contour, and so  there is no reason to expect that it should be a
physical quantity. Therefore, it is important to know how to extract
physical information from it, in the same way that the spectral
density, a physical quantity, may be obtained from the Wightman
function, even though the Wightman function itself is not physical. 
The one notable exception in the literature is \cite{Liao:2018uxa}.
There, the authors studied many body chaos in a weakly interacting 2D system of
fermions with quenched disorder and computed the Lyapunov exponent
both for the unregularized $\eta = 1/2$ case and the symmetrically
regularized one, $\eta = 0$ (in our notation). They indeed found
that the two results disagree, and pointed out the regulator dependence of the OTOC. The conclusion that they drew is that, in the model considered,  the only special feature of the symmetrically regularized OTOC is a particular cancellations of divergencies in the computation, but the physical meaning behind this correlator remained obscure.

Here we have performed a more thorough analysis showing the regulator dependence of the
OTOC for two paradigmatic models, a weakly coupled $\phi^4$ matrix boson (at any $N$) and the SYK model. By comparing to ordinary Schwinger-Keldysh theory, 
we provide a simple diagrammatic proof regarding the reason why the
choice of the contour affects the OTOC, although the Lyapunov spectrum
becomes contour independent for theories that stay massless/gapless
even at finite temperature.
This is particularly relevant for the SYK model, which has been
extensively studied over the last years. Its largest Lyaponuv exponent,
which saturates the MSS bound, is indeed contour independent. Similarly, in 2-dimensional CFTs in the limit of large central charge where the vacuum dominance emerges, the Lyapunov exponent is  insensitive to the choice of contour.

These detailed studies allow us to recognize that the regulator
dependence is an IR issue, and not an alleviation of purported UV
singularities. This means one has to take more care in understanding
the role of the regulator as it may contain physical information. 
One crucial insight of our paper is to recognise the special physical
meaning of the symmetrically regularised OTOC, by means of kinetic
theory \cite{Grozdanov:2018atb}. The OTOC computed on this
  contour is the one which one can properly claim to compute chaos or scrambling.  That the fact that the bound on chaos
holds for this physically meaningful definition of OTOC is remarkable
and open new directions on possible still unknown dynamical
constraints that the bound can impose.

This does then raise the question which simple observable naturally
gives rise to such a symmetric insertion of a thermal density matrix.
We proposed a simple observable, related to the operator fidelity,
which contains information beyond the commutator-squared and  can be
measured experimentally using echo spectroscopy.  The corollary of
using this observable to define the OTOC is that is based on a double
insertion of density matrices, i.e. the periodicity in Euclidean time
is twice the inverse temperature. From this point of view the bound on
chaos should read $\lambda \leq \pi k_B T_{\text{phys}}/\hbar$.

Overall, our results pose the question on the usefulness of the commutator-squared to probe quantum chaos. 
 The contour dependence of the commutator-squared and of the Lyapunov
 spectrum extracted from it casts doubts on whether the commutator
 square is physical and  how physical information should be extracted
 from it. However, even though a natural way to define chaotic quantum
 system is that in which the OTOC displays an exponential growth, this
 growth regime 
 actually clashes with the other notion of a quantum chaotic theory
 that it should display random matrix behaviour. In the SYK model,
 even though one has exponential growth at shorter times similar to
 {\it classical} weakly interacting chaos, spectral
 properties, such as the spectral form factor, are similar to that of
 random matrix theory for times of order of $N\log(N)$ and larger
 \cite{Cotler2016,Garcia-Garcia:2016mno,Garcia-Garcia:2017pzl}. This suggests that the model becomes truly {\it
   quantum} chaotic after this time-scale. A gorgeous example of
 true {\it quantum} chaos embodied by random matrix behaviour has been
 observed on the kicked Ising spin-$1/2$ chain for much shorter
 timescales \cite{Bertini2018a}. There is no exponential growth in the
 OTOC in this model, which challenges the notion of how quantum chaos
 and especially maximal chaos should be defined.

\acknowledgments
The authors have benefited from discussions with
Nikolay Gnezdilov,
Razvan Gurau,
Alexander Krikun, Tomaz Prosen,
Moshe Rozali,
Philippe Sabella-Garnier,
 and wish to thank Yingfei Gu and Subir Sachdev for emphasizing
  the contour independence of the SYK Lyapunov spectrum and correcting
  an earlier version.

This research was supported in part by a VICI award of the Netherlands
Organization for Scientific Research (NWO), by the Netherlands
Organization for Scientific Research/Ministry of Science and \mbox{Education}
(NWO/OCW), and by the Foundation for Research into Fundamental Matter (FOM).

\appendix\section{Numerical calculation in matrix model}\label{{app:matrix}}
In this appendix we outline the simplifications used to solve numerically the Bethe-Salpeter equation Eq. \eqref{eq:BS_eta}. Following \cite{Stanford2016}, we define
\beq
P = |\p|,\quad K = |\k|,\quad y = |\k - \p|
\eeq
and express the momentum integral as follows
\beq
\int d^3k = 2\pi \int_0^\infty K^2dK\int_{|K-P|}^{K+P} \frac{y dy}{KP}.
\eeq
Rewriting Eq. \eqref{eq:BS_eta} in the time domain and replacing the momentum integral, we arrive at the simplified version of the Bethe-Salpeter equation, which we solve numerically following the strategy described in \cite{Stanford2016}:
\begin{align}
\lambda_L f(P)= 
\int\limits_0^\infty {\dd K} 
&\left[
\left(
\cosh(\eta E_+)f(K)-
{f(P)\over3} {\sinh\left({\beta E_P\over2}\right)\over \sinh\left({\beta E_K\over2}\right)}\right) I_+(P,K)\right.\nonumber \\
&
\left.
+\left(
\cosh(\eta E_-)f(K)-
{f(P)\over3} {\sinh\left({\beta E_P\over2}\right)\over \sinh\left({\beta E_K\over2}\right)}\right) I_-(P,K)
\right]\,,\label{eq:Lyapunov_weak}
\end{align}
where
\begin{align}\label{eq:IKernel}
I_+(P,K) &\equiv  
{K\over (2\pi)^2 P 4 E_P E_K} 
\int\limits_{|K-P|}^{K+P} \dd y y R(E_+,y) =
 {{3\tilde{g} K\over (2\pi)^3 \beta P E_P E_K} \over \sinh( \beta E_+/2)}
 \int\limits_{|K-P|}^{K+P} \dd y
\log  {\sinh x_+^+\over \sinh x_-^+} \nonumber\\
 I_-(P,K) &\equiv  {K\over (2\pi)^2 P 4 E_P E_K} 
\int\limits_{|K-P|}^{K+P} \dd y y R(E_-,y) =
{ {3\tilde{g} K\over (2\pi)^3 \beta P E_P E_K} \over
\sinh( \beta E_-/2)}
 \int\limits_{|K-P|}^{K+P} \dd y
 \log {1-e^{-2x_+^-}\over 1-e^{2x_-^-}}\nonumber\\
 &x^+_\pm = \frac{\beta}{4}\left(E_+\pm y\sqrt{1 + \frac{4m^2}{y^2-E_+^2}}\right) \hspace{20pt} x^-_\pm = \frac{\beta}{4}\left(E_-\pm y\sqrt{1 + \frac{4m^2}{y^2-E_-^2}}\right)\,,
\end{align}
and we defined  $\tilde{g}=g^4(N^2+5)/(4\cdot 144)$.

\bibliographystyle{JHEP}
\bibliography{library}

\end{document}